\documentclass[pdftex,twocolumn,epjc3]{svjour3}  
\RequirePackage[T1]{fontenc}
\smartqed  
\RequirePackage{graphicx}
\RequirePackage{mathptmx}  
\RequirePackage{flushend}
\RequirePackage[numbers,sort&compress]{natbib}
\RequirePackage[colorlinks,citecolor=blue,urlcolor=blue,linkcolor=blue]{hyperref}
\usepackage{physics,amsmath}
\usepackage{relsize,exscale}
\usepackage{amssymb}
\usepackage{bm}
\usepackage{multirow}
\usepackage[caption=false]{subfig}
\usepackage[utf8]{inputenc}
\usepackage[utf8]{inputenc}
\usepackage[utf8]{inputenc}
\usepackage[utf8]{inputenc}
\journalname{Eur. Phys. J. C}
\begin{document}

\title{Luminosity calibration by means of van-der-Meer scan for Q-Gaussian beams}

\author{Mohamed A. Abed\thanksref{e1,addr1,addr2}
        \and
        Anton A. Babaev\thanksref{addr1}
        \and
        Leonid G. Sukhikh\thanksref{addr1}
}

\thankstext{e1}{e-mail: abedmohamed@tpu.ru (corresponding author)}

\institute{Tomsk Polytechnic University, Tomsk, Russia \label{addr1}
           \and
           Ain Shams University, Cairo, Egypt \label{addr2}
}

\date{Received: date / Accepted: date}

\maketitle

\begin{abstract}
	
	Luminosity is the key quantity characterizing the performance of charged particle colliders. Precise luminosity determination is an important task in collider physics. Part of this task is the proper calibration of detectors dedicated for luminosity measurements. The wide-used experimental method of calibration is the van-der-Meer scan, which is the beam separation scan performed at specifically optimized beam conditions. This work is devoted to modeling this scan with the q-Gaussian distribution of particles in colliding beams. Because of its properties, the Q-Gaussian distribution is believed to describe the density closer to reality than regular Gaussian-based models. In this work, the q-Gaussian model is applied for van-der-Meer scan modeling, and the benefits of this model for luminosity calibration task are demonstrated.

\end{abstract}

\section{Introduction}
	\label{intro}
	Colliders have developed immensely in the last decades; they have made a forefront contribution in exploring and discovering new physics \cite{r1}. The collider performance is defined by its luminosity $\mathcal{L}$ and the available energy in the centre of mass of colliding beams $E^{CoM}$, where the luminosity characterizes the intensity of particles collisions at the interaction point (IP) and the centre of mass energy shows the ability of the collider to produce heavier particles or probe smaller scales \cite{r2}. These two parameters are tuned in experiments to discover rare and new events. The probability of discovering new events can be increased in two ways: 1– Increase the event cross-section, which manifests the probability of a particular class of events to take place, which can be accomplished by increasing the energy of the colliding beams, which demands hard facility upgrades; 2– Increase the luminosity through the optimization of the collision conditions, which is more accessible.
	
	 Luminosity is often measured with dedicated detectors – luminometers. To perform precise measurements, these detectors must be calibrated properly. Several luminosity-calibration methods are used for hadron colliders and are summarized in \cite{r2}. A purely experimental technique based on the beam separation scan, which is the so-called van-der-Meer (vdM) scan, was successfully used at ISR \cite{r3,r4}, RHIC \cite{r5,r6,r7}, and LHC \cite{r8,r9,r10}. In a vdM scan, two beams are swept across each other, and the response of the luminometer under test is measured as a function of the distance between beam orbits (beam separation). The response rates are then fitted by Gaussian or double Gaussian fit models.
	
	Regularly, the models used to extract observables from vdM scan data imply a Gaussian distribution of particles in colliding beams, which restricts the precision of calibration. For the high-energy colliders, the actual particle densities deviate from the exact Gaussian. In \cite{r11,r12}, it was found that the colliding bunches have non-Gaussian tails. The non-Gaussianity of the tails can be attributed to the different effects they experience due to intra-beam effects such as intra-beam scattering, synchrotron radiation, quantum excitation, and due to the mutual interaction of the two colliding beams such as beam-beam effect, e-cloud, and luminosity burn-off. These combined effects change the tail-to-core population of the colliding bunches. Therefore, the effect of the non-Gaussian tails on the absolute luminosity should be investigated for precise luminosity calibration. Some artificial corrections (like a second Gaussian \cite{r10} or deformation of the scan curve with a polynomial \cite{r9}) are often introduced to take non-Gaussian tails into account. In \cite{r12,r13}, it was observed that the q-Gaussian distribution \cite{r14} provides a more realistic model for the actual bunch profile for LHC and the HL-LHC upgrade; therefore, it provides a more natural base study the effect of their non-Gaussian tails on the main luminosity parameters. In \cite{r15}, the impact of non-Gaussian tails on the absolute luminosity and emittance evolution was studied using a q-Gaussian model for bunch profile to investigate the modification of bunch shape due to the combined effects of intra-beam scattering and synchrotron radiation. 
	
	 The aim of this work is to estimate the influence of the non-Gaussian tail on the luminosity calibration by assuming q-Gaussian bunches, as well as to look for ways to obtain more precise vdM scan models. The work is structured as follows: In section \ref{sec:2}, the luminosity concept is introduced, and the main parameters of the vdM scan are discussed. In section \ref{sec:3}, a full description of the q-Gaussian distribution is given. After that, the analytical formulas for the overlap integral and convolved beam sizes for q-Gaussian bunches are derived, and their deviation from that of Gaussian bunches is estimated. Then, the effect of the tilt angle in the transverse plane of the colliding bunches is considered, and its effect on the overlap integral is estimated. In section \ref{sec:4}, the vdM scan performed with q-Gaussian bunches is considered. First, the ``{\em toy}'' vdM scan is modeled where the beam overlap is calculated based on the theory developed in section \ref{sec:3}, three different fit models (Gaussian, Double Gaussian and q-Gaussian) are applied to this ``{\em toy}'' dataset and the best fit is determined at these conditions when scan shape has non-Gaussian tails. Then, the real experimental response rate data (from LHC CMS vdM program \cite{r10}) is tested to check if q-Gaussian fit model really works better than current Gaussian-based models.

\section{Luminosity concept and Van-der-Meer scan}
\label{sec:2}
	\subsection{Luminosity and overlap integral}
	\label{subsec:2.1}
	The luminosity depends only on the properties of colliding beams. The luminosity of single bunch-crossing is defined as \cite{r16}:
	\begin{equation}
		\mathcal{L}(\bm{\Delta}_r,t)=N_{1} N_{2} K\int\rho_1^{lab}(\vectorbold{r}-\bm{\Delta}_r,t)\rho_2^{lab}(\vectorbold{r},t)\,d^3\vectorbold{r}\,dt,
		\label{eq:1}
	\end{equation}
	where $N_{1,2}$ are the number of particles in the colliding\linebreak[4] bunches, which move with velocity  $\vectorbold{v}_1$ and $\vectorbold{v}_2$ correspondingly, $K$ is  Moller kinematic relativistic factor  $K=$\linebreak[4]$\sqrt{(\vectorbold{v}_1-\vectorbold{v}_2)^2-\frac{(\vectorbold{v}_1\times \vectorbold{v}_2)^2}{c^2}}$ \cite{r17}, $\rho_{1,2}^{lab}(\vectorbold{r},t)$ are the normalized particle distribution densities in the colliding bunches in the lab frame and $\bm{\Delta}_r$  is the separation between their centres. The multiplication of the Moller kinematic relativistic factor and the integral in equation \eqref{eq:1} represents the reciprocal of the effective area of the luminous region at the interaction point,\ $\Omega=1/A_{eff}$, which is the so-called overlap integral \cite{r18}. And its invariance form  can be obtained from \cite{r16,r19} as:
	\begin{equation}
		\Omega(\bm{\Delta}_{r_{\bot}})=\frac{1}{\gamma_\bot}\int\rho_1^{lab,\bot}(\vectorbold{r}_\bot-\bm{\Delta}_{r_{\bot}})\rho_2^{lab,\bot}(\vectorbold{r}_\bot)\,d\vectorbold{r}_\bot,
		\label{eq:2}
	\end{equation}
	where $\rho_{1,2}^{lab,\bot}(\vectorbold{r}_\bot)$ are the spatial transverse particle distributions of the colliding bunches in the lab frame, $\bm{\Delta}_{r_\bot}$  is the transverse beam separation and $\gamma_\bot$  is the Lorentz relativistic factor due to transverse boost.

	\subsection{Van-der-meer scan}
	\label{subsec:2.2}
	In principle, luminosity is measured using detectors reacting on the flux of collision products. They can be designed based on different effects and principles \cite{r10}. In this paper we are not interested in the specific process used for luminosity measurements, the measured quantity is called ``{\em response rates}'' and notated with $R$. 
	
	The van-der-Meer (vdM) scan was proposed by S. Van Der Meer \cite{r4}. It is based on separating two beams across each other in the transverse plane, while the response rate $R$  of their interaction is monitored as a function of the transverse separation distance between their orbits $\Delta$. If the colliding bunches have factorizable densities, $\rho^{lab}(\vectorbold{r}_\bot)=\rho_x(x)$\linebreak[4]$\rho_y(y)$, or in other words, if their overlap integral is factorizable, $\Omega(\bm{\Delta}_{r_\bot})=\Omega_x(\Delta_x)\ \Omega_y(\Delta_y)$, this implies that response rate  $R$  has no correlation between separations in the horizontal ``$x$'' and vertical ``$y$'' directions, thus two one-dimensional vdM scans are performed separately in each direction, and the convolved bean sizes (i.e. RMS widths of the resulting scan curves) are found as:
	\begin{equation}		
		\Sigma_x=\frac{\int R(\Delta_x,0)\, d\Delta_x}{C_x\, R(0,0)};\quad
		\Sigma_y=\frac{\int R(0,\Delta_y)\, d\Delta_y}{C_y\, R(0,0)},
		\label{eq:4}	
	\end{equation}
	where $C_{x,y}$  are constants and depend on the particle densities or, more precisely, the scan curve shape in the horizontal and vertical directions, and the maximum overlap integral and luminosity are given by 
	\begin{equation}		
		\Omega(0,0)=\frac{1}{C_xC_y\Sigma_x\Sigma_y};\quad
		\mathcal{L}(0,0)=fN_1N_2\Omega(0,0),
		\label{eq:5}
	\end{equation}
	consequently, the calibration constant, which is the so-called visible cross-section $\sigma^{vis}$, that relates the measurable response rates $R$ of a luminometer recorded by a given luminosity algorithm to the absolute luminosity $\mathcal{L}$, is determined as
	\begin{equation}		
		\sigma^{vis}=\frac{R(0,0)}{fN_1N_2\Omega(0,0)}.
		\label{eq:6}	
	\end{equation}
	where the visible cross-section $\sigma^{vis}$ is the ratio of measured rates to corresponding beam overlap. If the particle densities are not factorizable or, more specifically, their overlap integral has $x-y$ coupling, and the vdM scan is not precise.

	 In practice, the vdM scan is performed under special conditions where the beam parameters are optimized to determine calibration constant with high precision. The vdM scan formalism is valid for arbitrary particle densities and crossing angles \cite{r16}.
	 
\section{Theory}
 \label{sec:3}
	\subsection{Q-Gaussian probability distribution function}
	  \label{subsec:3.1}
	  The q-Gaussian distribution has diverse applications in generalized statistical theory, laser, plasma, and astronomy, see \cite{r20,r21,r22,r23,r24}, and it has been used to investigate emittance evolution and beam profile modeling for LHC \cite{r13,r15}. The q-Gaussian distribution is known for its remarkable  ability to represent a variety of distributions, from bounded finite distributions such as rectangular distribution at ``$q\to-\infty$ ” and parabolic distribution at ``$q\to0$ ” to heavy-tailed infinite distributions such as Student’s t-distribution, where $q$ controls the tails population.
	 The q-Gaussian distribution is shown in Fig. \ref{subfig:2a} and is defined as:
	 \begin{equation}		
	  QG(u;q,\beta^{qG})=\frac{\sqrt{\beta^{qG}}}{C^{qG}}e_q(-\beta^{qG}u^2),
	 \label{eq:8}	
	 \end{equation}
	 where $\beta^{qG}$ is a real positive number, $e_q$  is q-exponential and  $C^{qG}$ is the normalization constant, and they are defined as:
	\begin{equation}		
		e_q(-\beta^{qG}u^2)=
		\begin{cases}
			\exp(-\beta^{qG}u^2) & \text{if } q = 1\\
			[1-(1-q)\beta^{qG}x^2]_{+}^{\frac{1}{1-q}} & \text{if } q \ne 1\\
		\end{cases},	
		\label{eq:9}
	\end{equation}
	\noindent and
	\begin{equation}		
		C^{qG}=
		\begin{cases}
			\frac{2}{(3-q)\sqrt{1-q}} Beta(\frac{1}{1-q},\frac{1}{2}) & \text{if } q < 1\\
			\sqrt{\pi} & \text{if } q = 1\\
			\frac{1}{\sqrt{q-1}} Beta(\frac{1}{q-1}-\frac{1}{2},\frac{1}{2}) & \text{if } 1<q<3
		\end{cases}.	
		\label{eq:10}
	\end{equation}
	 For $q<1$, the core is blown up, and the distribution becomes finite with light tails with $u\in\left[-\frac{1}{\sqrt{(1-q)\beta^{qG}}},\frac{1}{\sqrt{(1-q)\beta^{qG}}}\right]$. For $q>1$,  the tail density increases, and the distribution becomes heavy-tailed. At $q=1$,  the standard Gaussian distribution is restored. The standard deviation of the q-Gaussian distribution is dependent on $q$  and $\beta^{qG}$, and is given by
	\begin{equation}		
	 	\sigma^{qG}=
	 	\begin{cases}
	 		\frac{1}{\sqrt{(5-3q)\beta^{qG}}}  & \text{if } q < \frac{5}{3}\\
	 		\infty & \text{if } \frac{5}{3}\le q < 2\\
	 		Undefined & \text{if } 2 \le q<3
	 	\end{cases}.
	 	\label{eq:11}
	 \end{equation}
	 The standard deviation $\sigma^{qG}$ represents the RMS bunch dimension; therefore, in the q-Gaussian bunch model, the range of the tail weight $q$ is limited to $q<\frac{5}{3}$. 
	 
	\subsection{Overlap integral of q-Gaussian bunches}
	\label{subsec:3.2}
	Let’s assume two bunches, 1 and 2, with transverse particle densities $\rho_{1,2}(x,y)$  in the lab frame, the two bunches collide head-on (i.e. there is no crossing angle $\phi=0$) and the bunches have factorizable particle densities in horizontal and vertical directions. If the two bunches are separated in opposite directions with separation  $\frac{\Delta_u}{2}$, the one-dimensional overlap integral $\Omega_u$ $(u=x,y)$ can be written as:
	\begin{equation}		
		\Omega_u(\Delta_u)=\int\rho_1 \left( u-\frac{\Delta_u}{2}\right) \rho_2\left(u+\frac{\Delta_u}{2}\right)\, du.
		\label{eq:12}
	\end{equation}
	If the two bunches have q-Gaussian particle densities, and they have equal dimensions and tail densities in their respective direction (i.e. $\sigma_{1u}=\sigma_{2u}=\sigma_{u}$ and $q_{1u}=q_{2u}=q$ ), equation \eqref{eq:12} becomes
	\begin{align}		
		\Omega_u^{qG}(\Delta_u;q)=&\frac{\beta^{qG}}{{C^{qG}}^2} \mathop{\mathlarger{\mathlarger{\int}}} e_q \left(-\beta^{qG} \left(u-\frac{\Delta_u}{2}\right)^2\right)\nonumber \\
	  	&\times e_q \left(-\beta^{qG} \left(u+\frac{\Delta_u}{2}\right)^2\right)\,du,
		\label{eq:13}
	\end{align}
	where $\beta^{qG}$ is determined from equation \eqref{eq:11}. For $q=1$, the q-Gaussian is equivalent to the normal Gaussian, and  $\Omega_u^{qG}(\Delta_u,1)$ is given by
	\begin{equation}		
		\Omega_u^{qG}(\Delta_u;1)=\frac{\sqrt{\beta^{qG}}}{\sqrt{2}\, C^{qG}} \exp \left(-\beta^{qG}\frac{\Delta_u^2}{2}\right).
		\label{eq:14}
	\end{equation}
	For $q\ne 1$, using $e_q$  definition from equation \eqref{eq:9} can be written as:
	\begin{align}
		\Omega_u^{qG}&\left(\Delta_u;q\ne1\right)\nonumber\\
		 &=\frac{\beta^{qG}}{{C^{qG}}^2} \mathop{\mathlarger{\mathlarger{\int}}} \left[1-(1-q)\beta^{qG}\left(u-\frac{\Delta_u}{2}\right)^2\right]_{+}^{\frac{1}{1-q}}\nonumber\\
		&\quad \times \left[1-(1-q)\beta^{qG}\left(u+\frac{\Delta_u}{2}\right)^2\right]_{+}^{\frac{1}{1-q}}\, du.
		\label{eq:15}
	\end{align}
	For finite light-tailed bunches with $q<1$, equation \eqref{eq:15} becomes:
	\begin{align}		
		\Omega_u^{qG}&\left(\Delta_u;q<1\right)\nonumber\\
		&=\frac{\beta^{qG}}{{C^{qG}}^2} 	\mathop{\mathlarger{\mathlarger{\int}}}_{u_1}^{u_2} \left[1-(1-q)\beta^{qG}\left(u-\frac{\Delta_u}{2}\right)^2\right]_{+}^{\frac{1}{1-q}}\nonumber\\
		&\quad \times 	\left[1-(1-q)\beta^{qG}\left(u+\frac{\Delta_u}{2}\right)^2\right]_{+}^{\frac{1}{1-q}}\, du.
		\label{eq:16}
	\end{align}	
	Since the bunches have finite tails, the overlap integral is finite over the region where the densities of the two bunches can overlap, hence the integration limits $u_{1}$  and $u_{2}$  are given by 
	\begin{align*}	
		&\{u_1,u_2\}\nonumber
		\\&\quad \quad =
		\begin{cases}
			\bigg\{-\frac{1}{\sqrt{(1-q)\beta^{qG}}} -\frac{\Delta_u}{2},\frac{1}{\sqrt{(1-q)\beta^{qG}}} +\frac{\Delta_u}{2}\bigg\}, \\
			 \hfill \text{if } -\frac{2}{\sqrt{(1-q)\beta^{qG}}} < \Delta_u <0\vspace{0.5 mm} \\ 
			\bigg\{-\frac{1}{\sqrt{(1-q)\beta^{qG}}} +\frac{\Delta_u}{2},\frac{1}{\sqrt{(1-q)\beta^{qG}}} -\frac{\Delta_u}{2}\bigg\}, \\
			 \hfill \text{if } 0 \le \Delta_u < \frac{2}{\sqrt{(1-q)\beta^{qG}}}
		\end{cases}.
	\end{align*}
	By solving equation \eqref{eq:16}, The general form of the overlap integral of light-tailed q-Gaussian bunches with equal bunch sizes and tail densities $\Omega_u^{qG}\left(\Delta_u;q<1\right)$ is obtained as:
	\begin{align}
			 &\Omega_u^{qG}\left (\Delta_u;q<1\right)=\frac{\sqrt{\beta^{qG}}}{\sqrt{1-q}\ {C^{qG}}^2} \left(1-\sqrt{(1-q)\beta^{qG}\frac{\Delta_u^2}{4}} \right)\nonumber\\ 
			 &\times \left(1-{(1-q)\beta^{qG}\frac{\Delta_u^2}{4}}  \right)^\frac{2}{1-q} Beta\left(\frac{1}{2},\frac{2-q}{1-q} \right)\nonumber \\ 
			 &\times \ _2F_1\left(\frac{-1}{1-q},\frac{1}{2};\frac{5-3q}{2-2q};\left( \frac{1-\sqrt{(1-q)\beta^{qG} \Delta_u^2/4}}{1+\sqrt{(1-q)\beta^{qG} \Delta_u^2/4}} \right)^2\right), 
		\label{eq:17}
		\end{align}
	where$\ _2F_1$ is the Gaussian hypergeometric function \cite{r25}. For a detailed derivation of equation \eqref{eq:17}, see \ref{app:A}.
	
	For infinite heavy-tailed bunches with $q>1$ , equation \eqref{eq:15} becomes:
	\begin{align}
			\Omega_u^{qG}&\left(\Delta_u;q>1\right)\nonumber\\
			&=\frac{\beta^{qG}}{{C^{qG}}^2} 	\mathop{\mathlarger{\mathlarger{\int}}}_{-\infty}^{\, \infty} \left(1-(1-q)\beta^{qG}\left(u-\frac{\Delta_u}{2}\right)^2\right)^{\frac{1}{1-q}}\nonumber\\
			\quad &\times \left(1-(1-q)\beta^{qG}\left(u+\frac{\Delta_u}{2}\right)^2\right)^{\frac{1}{1-q}} \,du,
			\label{eq:18}
	\end{align}
	by solving equation \eqref{eq:18}, The general form of the overlap integral of heavy-tailed q-Gaussian bunches with equal bunch sizes and tail densities  $\Omega_u^{qG}\left (\Delta_u;q>1\right)$ is obtained as:
	\begin{align}
			\Omega_u^{qG}&\left (\Delta_u;q>1\right)=\frac{\sqrt{\beta^{qG}}}{\sqrt{q-1}\ {C^{qG}}^2} Beta\left( \frac{1}{2},\frac{5-q}{2q-2}\right)\nonumber\\ 
			&\times\ _2F_1\left(\frac{1}{q-1},\frac{5-q}{2q-2};\frac{q+1}{2q-2};(1-q)\beta^{qG}\frac{\Delta_u^{2}}{4} \right). 
			\label{eq:19}
		\end{align}
	 For a detailed derivation of equation\eqref{eq:19}, see \ref{app:A}.
	
	 The crossing angle effect can be considered in equations \eqref{eq:14}, \eqref{eq:17} and \eqref{eq:19} through the modification of the transverse bunch size as in \cite{r19}.
	 
	 At the limit $q$ tends to 1, the q-Gaussian bunches tend to Gaussian; therefore, the overlap integral of the q-Gaussian bunches should have the same tendency. Thus, the limit of equations \eqref{eq:17} and \eqref{eq:19}  is evaluated as in equation \eqref{eq:20}, and it shows the fulfillment of the tendency. See \ref{app:B} for details.
	 \begin{align}
	 	\lim_{q \to 1}\left( \Omega_u^{qG}\left(\Delta_u;q<1\right) \right)&=\lim_{q \to 1}\left( \Omega_u^{qG}\left(\Delta_u;q>1\right) \right)\nonumber\\&
	 	=\Omega_u^{qG}\left(\Delta_u;q=1\right). 
	 	\label{eq:20}
	 \end{align}
 
 	\subsection{Difference between regular Gaussian and Q-Gaussian beams of equal RMS beam size }
 	\label{subsec:3.2*}
\begin{figure}
		\begin{center}
		\subfloat[\label{subfig:2a}]{\includegraphics[width=0.9\linewidth]{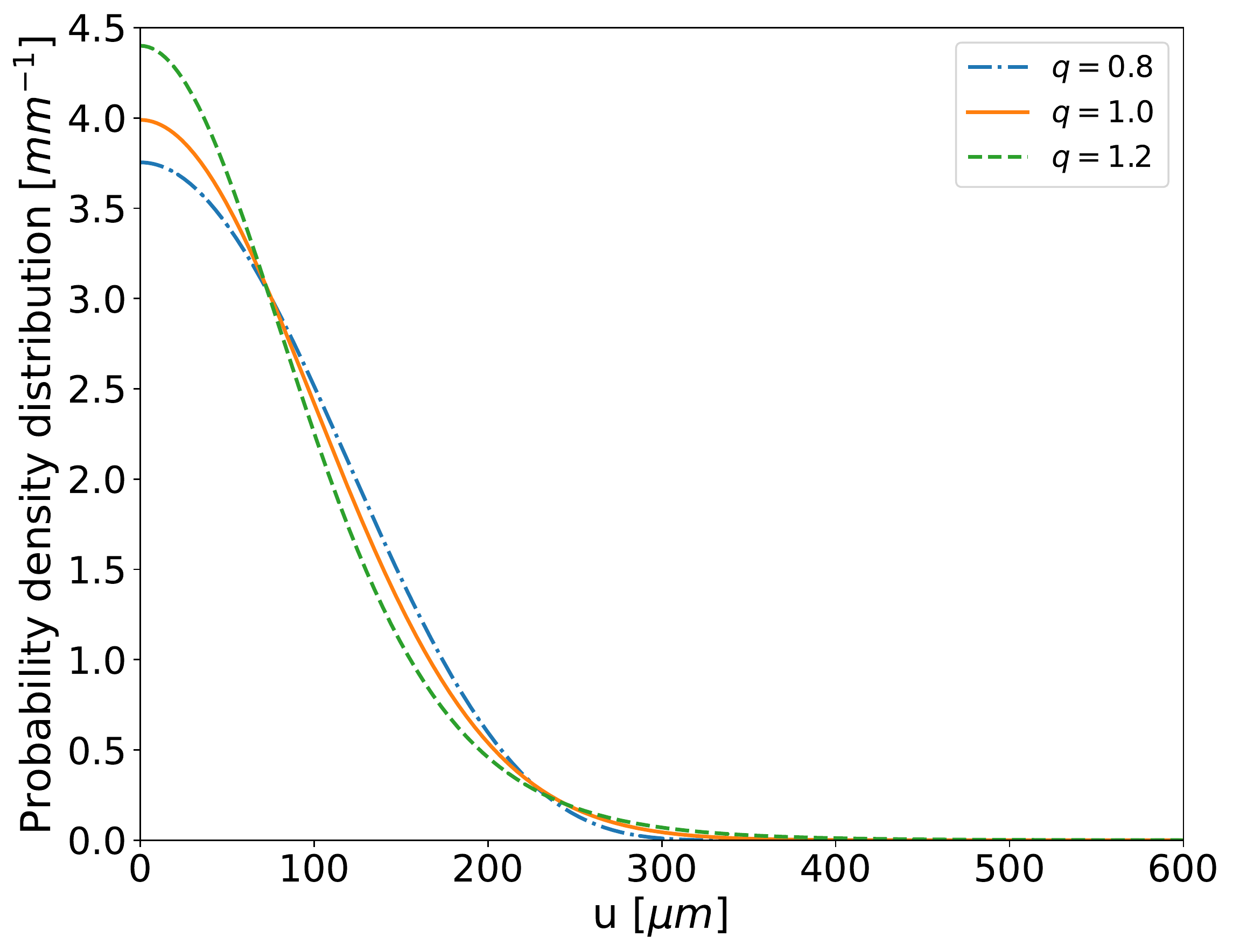}}\\
		\subfloat[\label{subfig:2b}]{\includegraphics[width=0.9\linewidth]{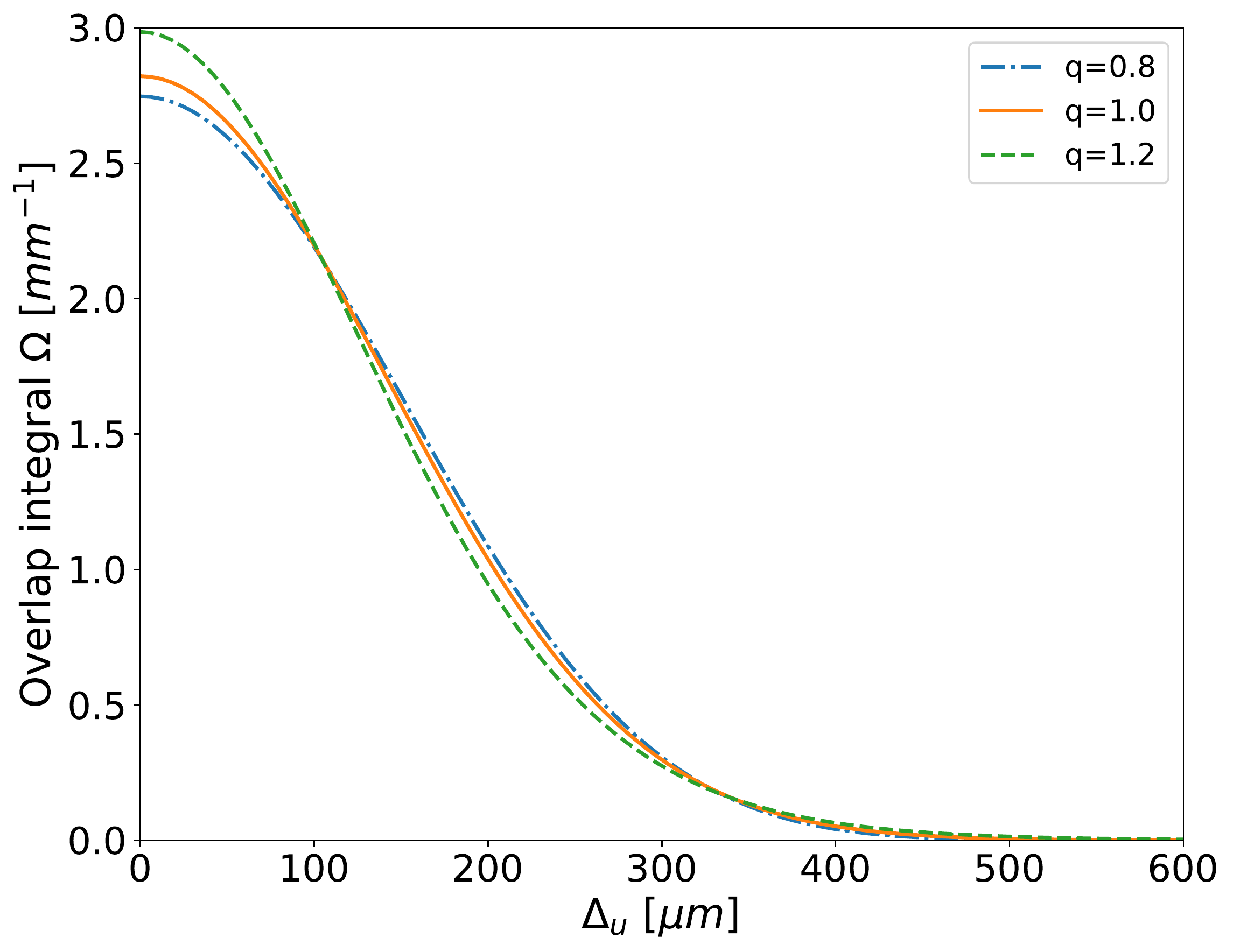}}
		\end{center}
		\caption{The bunch profile of q-Gaussian bunches with dimension  $\sigma_{u}^{qG}=100\ \mu m$ and tail densities $q=0.8$, $1.0$ and $1.2$ (\ref{subfig:2a}) and their overlap integral during the separation scan over separation $\Delta_u$ in the range $0$ to $6$ $\sigma_u^{qG}$ (\ref{subfig:2b})}
	    \label{fig:2}
	\end{figure}

\begin{figure}
		\begin{center}
			\includegraphics[width=0.9\linewidth]{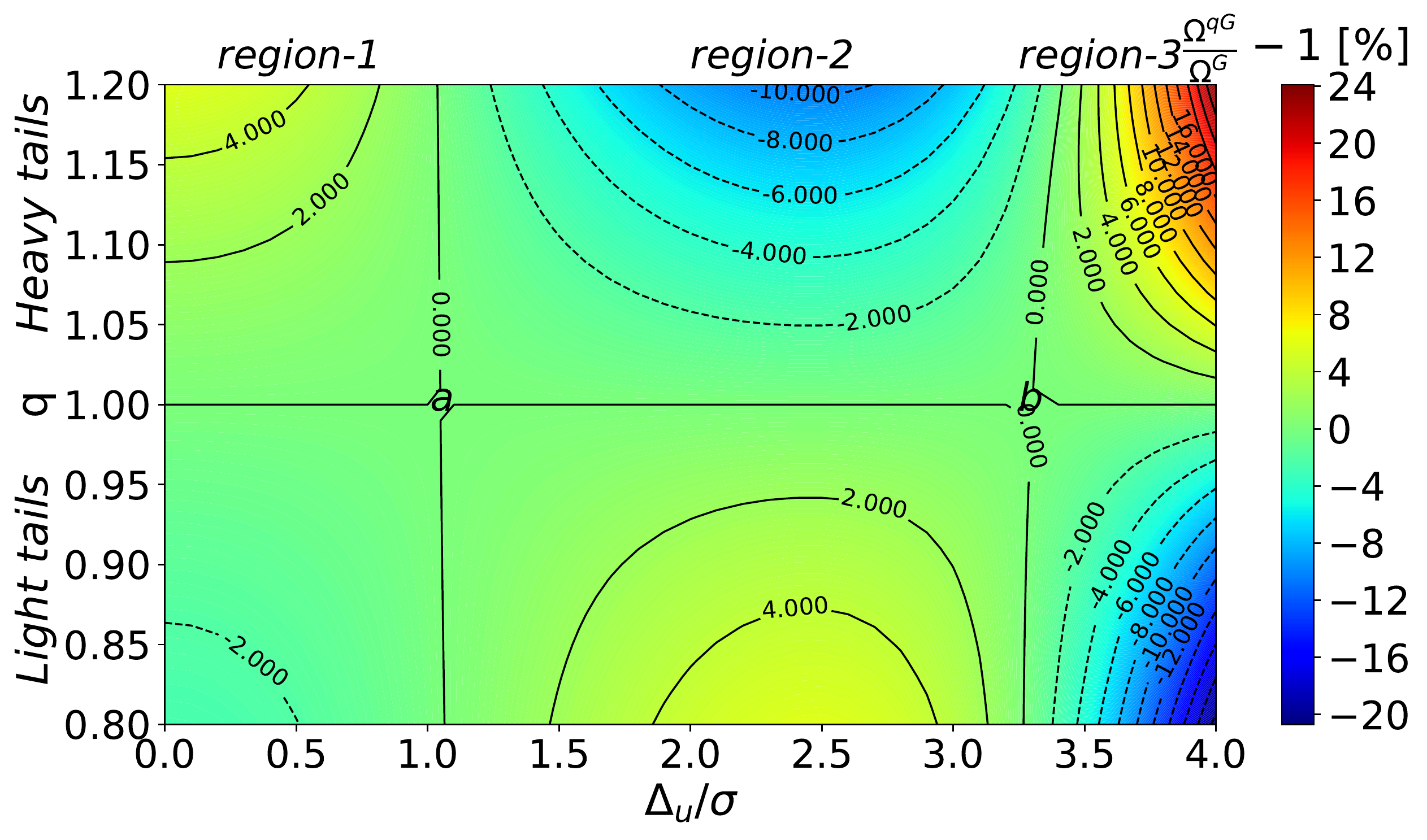}
		\end{center}
		\caption{The deviation map of the one-dimensional overlap integral of q-Gaussian bunches $\Omega_u^{qG} $ from that of the Gaussian $\Omega_u^{G}$  for bunches with equal dimensions $\sigma_{u}^{qG}=100\ \mu m$  with tail densities $q$ in the range $0.8$ to $1.2$ for vdM separation scan with separation $\Delta_u$ in the range $0$ to $4$ $\sigma_u^{qG}$}
		\label{fig:3}
	\end{figure}

	 Since the actual particle densities in the colliding bunches have non-Gaussian tails, the impact of the non-Gaussian tails on the overlap integral should be investigated. In order to do so, the separation scan for q-Gaussian bunches is modeled by using the analytical formula for the overlap integral $\Omega^{qG}$  in equations \eqref{eq:17} and \eqref{eq:19} with bunch parameters from van-der-Meer scans performed at CMS experiment (CERN). The vdM scan special conditions are $\sigma_u=100\ \mu m$  with no crossing angle \cite{r10}. In the investigation. Three beams of the same RMS beam size $\sigma_u^{qG}=100\ \mu m$ and different tail densities $q=0.8$, $1.0$ and $1.2$ were considered to represent bunches with light tails, Gaussian, and heavy tails, respectively. The bunch profiles are shown in Fig. \ref{subfig:2a}, and their respective overlap integral $\Omega^{qG}$  over separation $\Delta_u$  in the range of $0$ to $6$ $\sigma_u^{qG}$ are shown in Fig. \ref{subfig:2b}. At zero separation ($\Delta_u=0$), the heavy-tailed bunches have the highest overlap integral. Remarkable deviations of overlap integral of the q-Gaussian bunches  $\Omega^{qG}$  from that of Gaussian  $\Omega^{G}$  are also observed in Fig. \ref{subfig:2b}; these deviations are dependent on the separation. Thus to estimate this dependency, a deviation map was constructed, as shown in Fig. \ref{fig:3}, where a beam separation scan with the separation $\Delta_u$  in the range $0$ to $4$ $\sigma_u^{qG}$  is modeled, and the deviation of the overlap integral of q-Gaussian bunches to the Gaussian with identical bunch dimensions is estimated for different tail densities $q$ in the range $0.8$ to $1.2$. This tail density range corresponds to a tail population that differs from Gaussian by up to $20\%$.
	
	 The deviation map shows that the dependency of deviation on the separation $\Delta_u$  is divided into 3 regions: region 1– form zero separation to point ``{\em a\/}'', the overlap integral is higher for heavy-tailed beams (lower for light-tailed beams), and it decreases (increases) as the separation increase until it equals to the overlap integral of Gaussian beams at point ``{\em a\/}''; region 2– from point ``{\em a\/}'' to point ``{\em b\/}'', the overlap integral is lower for heavy-tailed beams (higher for light-tailed beams), and it decreases (increases) as the separation increases until it reaches a minimum (maximum), then, it increases (decreases) with further separation until it equals to that of the Gaussian at point ``{\em b\/}''; region 3– from point ``{\em b\/}'', the overlap integral is higher for heavy-tailed beams (lower for light-tailed beams), and it increases (decreases) as the separation increases. The limits of these deviations are summarized in Table \ref{tab:1}. The positions of point ``{\em a\/}'' and point ``{\em b\/}'' are different for different tail densities $q$ values, and they fall in the range $1.037$ to $1.065$ $\sigma_u^{qG}$ and $3.269$ to $3.416$ $\sigma_u^{qG}$, respectively. This sensitivity of the overlap integral to the tail density of the colliding bunches justifies the need for precise consideration of non-Gaussian beam shape in luminosity modeling. 
	\begin{table}
		\centering
		\caption{The deviation limits of the overlap integral of the one-dimensional overlap integral of q-Gaussian bunches $\Omega_u^{qG} $ from that of the Gaussian $\Omega_u^{G}$ at different dependency regions for the vdM separation scan with separation $\Delta_u$ in the range $0$ to $4$ $\sigma_u^{qG}$   for bunches with equal dimensions $\sigma_{u}^{qG}=100\ \mu m$  with tail density $q$  in the range $0.8$ to $1.2$}
		\label{tab:1}      
		\begin{tabular*}{\columnwidth}{@{\extracolsep{\fill}}lrrr@{}}
			\hline
			Tail population density & Region-1  & Region-2 & Region-3	\\
			\hline
			Light-tailed $q<1$		& $-2.64\%$ & $5.75\%$ & $-20\%$	\\
			Heavy-tailed $q>1$		& $5.79\%$	& $-10.46\%$ & $24\%$	\\
			\hline
		\end{tabular*}
	\end{table}

	\subsection{Convolved beam size of q-Gaussian bunches}
	\label{subsec:3.3}
	 Since the response rate $R$ is proportional to overlap integral $\Omega$, the one-dimensional convolved beam size of two q-Gaussian bunches $\Sigma_u^{qG}$ is found from equation \eqref{eq:4} as
	 \begin{equation}
	 	\Sigma_u^{qG}(q)=\frac{\int \Omega_u^{qG}(\Delta_u;q)\,d \Delta_u}{C_u\ \Omega_u^{qG}(0;q)},
	 	\label{eq:21}
	 \end{equation} 
	where the constant $C_u$ is taken as $\sqrt{(5-3q)\ C^{qG}}$ for \linebreak[4] q-Gaussian bunches. Since for any arbitrary normalized particle densities $\int \Omega_u(\Delta_u) \,d\Delta_u=1$ \cite{r16}, therefore for $\sigma_{1u}=\sigma_{2u}=\sigma_u$ and $q_{1u}=q_{1u}=q$, the convolved beam size $\Sigma_u^{qG}(q)$ is obtained as:
	 \begin{equation}
		\Sigma_u^{qG}(q)=\begin{cases}	 \frac{Beta\left(\frac{1}{2},\frac{3-q}{2q-2}\right)}{Beta\left(\frac{1}{2},\frac{5-q}{2q-2}\right)}\ \sigma_u^{qG}	
		& \text{if } 1<q<3 \\
		\sqrt{2}\ \sigma_u^{qG}	
		& \text{if } q=1 \\ \frac{Beta\left(\frac{1}{2},\frac{2-q}{1-q}\right)}{Beta\left(\frac{1}{2},\frac{3-q}{1-q}\right)}\ \sigma_u^{qG}	
		& \text{if } q<1 
	 	\end{cases}.
	  	\label{eq:22}
	 \end{equation}
	
	Similarly, the convolved beam size of q-Gaussian bunches with different bunch dimensions $\sigma_{1u}\ne\sigma_{2u}$ is presented in \ref{app:C}.
	
	Similar to the overlap integral, the convolved beam size of q-Gaussian bunches tends to that of Gaussian at the limit of $q$  tends to $1$.
	\begin{equation}
		\lim_{q \to 1}\left( \Sigma_u^{qG}(q<1)\right)=\lim_{q \to 1}\left(\Sigma_u^{qG}(q>1)\right)=\sqrt{2}\ \sigma_u^{qG}.
		\label{eq:23}
	\end{equation}
	
\begin{figure}
		\begin{center}
			\subfloat[\label{subfig:4a}]{\includegraphics[width=0.9\linewidth]{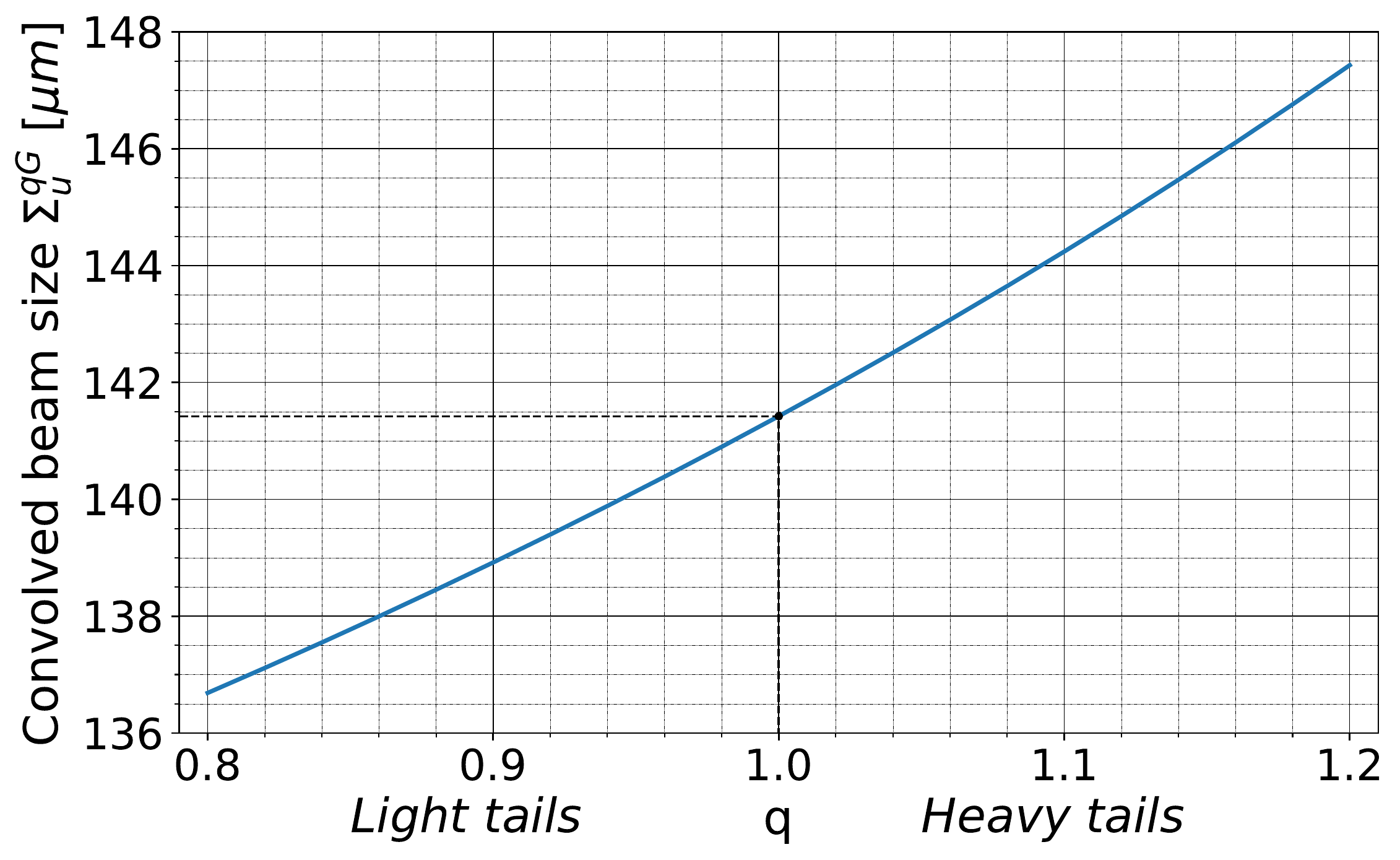}}\\
			\subfloat[\label{subfig:4b}]{\includegraphics[width=0.9\linewidth]{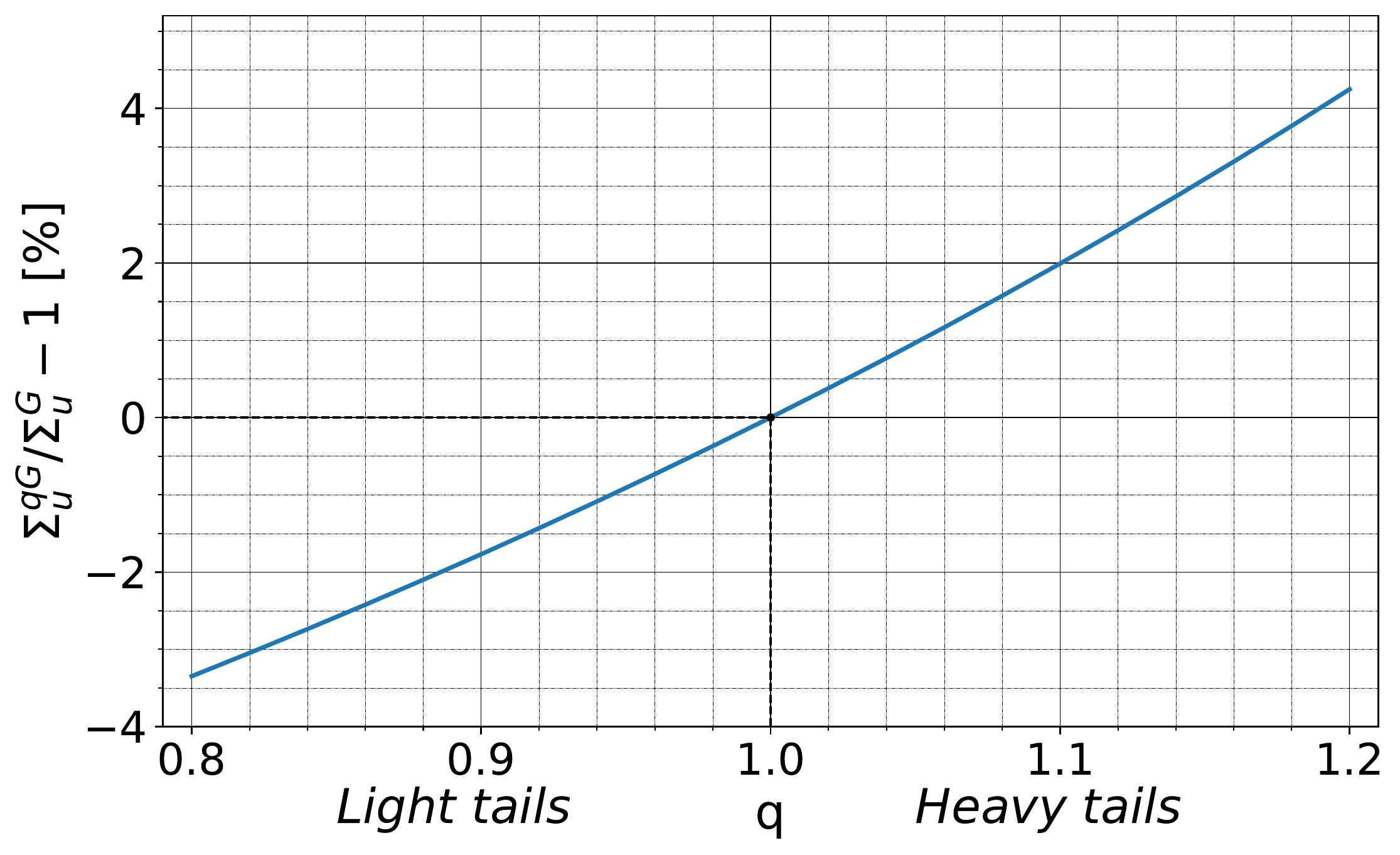}}
		\end{center}
		\caption{The convolved beam size of q-Gaussian bunches $\Sigma_u^{qG}$ for bunches with equal dimensions $\sigma_u^{qG}=100$ $\mu m$ and tail densities $q$ in the range $0.8$ to $1.2$ collide head-on (\ref{subfig:4a}), and its deviation from that of Gaussian bunches (\ref{subfig:4b})}
		\label{fig:4}
	\end{figure}

	Since, for precise luminosity calibration, the convolved beam size should be precisely defined from vdM scan curve; therefore it is essential to investigate the effect of\linebreak[4] non-Gaussian tail populations on the convolved beam size. The dependence of the one-dimensional convolved beam size of q-Gaussian bunches $\Sigma_u^{qG}$ on the tail density $q$ is modeled, using equation \eqref{eq:22}, for bunches with vdM scan special conditions of  $\sigma_u^{qG}=100$ $\mu m$ with no crossing angle, see \cite{r10}. The tail density $q$ in the range $0.8$ to $1.2$ is investigated. Figure \ref{subfig:4a} shows that for a certain bunch dimension, the convolved beam size increases as the tail density increases. Figure \ref{subfig:4b} shows the deviation of convolved beam size of q-Gaussian bunches $\Sigma_u^{qG}$ from that of Gaussian $\Sigma_u^{G}$, it was found that a difference of $20\%$ in tails population from Gaussian leads to a devotion up to $4.25\%$ for heavy-tailed bunches and down to $-3.35\%$ for light-tailed bunches.

	\subsection{The impact of the tilt angle}
	\label{subsec:3.4}
	When two bunches collide, their overlap integral depends on their transverse particle distributions in the lab frame, as in equation \eqref{eq:2}, In general, their respective horizontal and vertical axes in the transverse planes do not usually coincide, which leads to a small angle between the bunches transverse planes, which is the so-called tilt angle. The tilt angle is usually small, and it leads to transverse densities that have $x-y$ coupling in the lab frame as shown in Fig. \ref{fig:5}. Thus, the effect of the tilt angle on the overlap integral is investigated for bunches with Gaussian and non-Gaussian tails, and its impact on the vdM scan is considered. 
	
		\subsubsection{Gaussian tails}
		\label{subsubsec:3.4.1}
		Let’s assume two bunches with transverse densities $\rho_{1,2}$ are colliding at zero crossing angle, where the bunches have equal dimensions in their respective horizontal and vertical directions such that ($\sigma_{1x}=\sigma_{2x}=\sigma_x$ and $\sigma_{1y}=\sigma_{2y}=\sigma_y$), let's define three frames of reference: bunch 1 frame of reference $X'Y'Z'$; bunch 2 frame of reference $X''Y''Z''$ and lab frame of reference $XYZ$.
		\begin{figure}
		 	\begin{center}
		 		\subfloat[\label{subfig:5a}]{\includegraphics[width=0.5\linewidth]{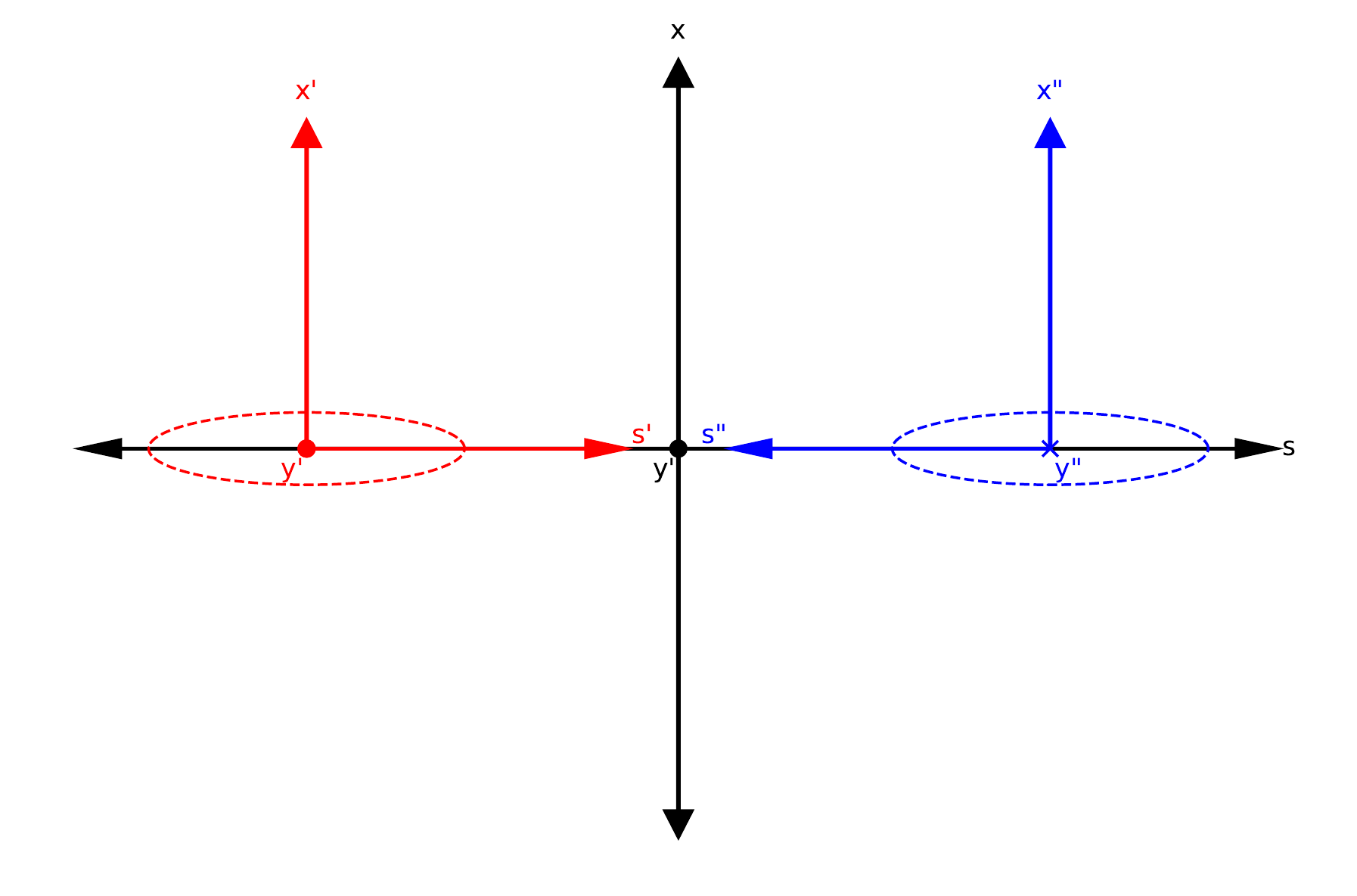}}	\subfloat[\label{subfig:5b}]{\includegraphics[width=0.5\linewidth]{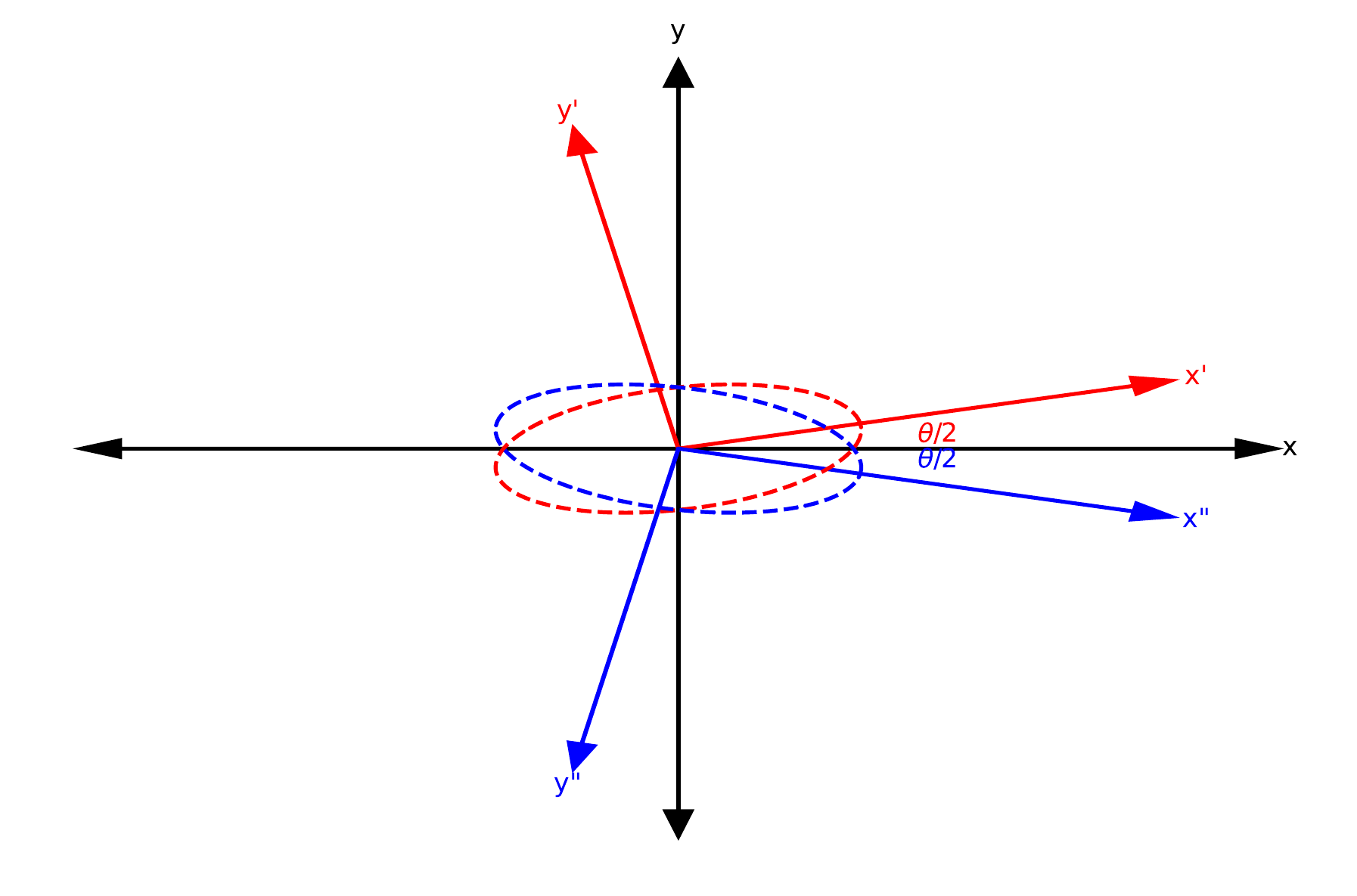}}
		 	\end{center}
		 	\caption{The Accelerator (\ref{subfig:5a}) and transverse (\ref{subfig:5b}) planes for bunch 1 (red) frame of reference $X'Y'Z'$; bunch 2 (blue) frame of reference $X''Y''Z''$ and lab frame of reference $XYZ$, where the transverse planes of bunch 1 and bunch 2 are rotated in opposite directions with a tilt angle $\theta$}
		 	\label{fig:5}
		\end{figure}
		In case of a tilt angle $\theta$ between $X'Y'$and $X''Y''$ such that the $X'Y'$ and $X''Y''$ are rotated with an angle $\frac{\theta}{2}$ in opposite directions around $XY$, as shown in Fig. \ref{fig:5}, the bunches coordinates are defined in the lab frame as:
		\begin{equation}
			\begin{split}
				x'&=x\cos \left( \frac{\theta}{2} \right) + y\sin \left(\frac{\theta}{2}\right);\\
				y'&= - x\sin \left( \frac{\theta}{2} \right) + y\cos \left(\frac{\theta}{2}\right);\\
				x''&=x\cos \left( \frac{\theta}{2} \right) - y\sin \left(\frac{\theta}{2}\right);\\
				y''&=- x\sin \left( \frac{\theta}{2} \right) - y\cos \left(\frac{\theta}{2}\right).
			\end{split}
			\label{eq:24}
		\end{equation}
		If there are horizontal and vertical separations $\Delta_x$ and $\Delta_y$, respectively, the overlap integral $\Omega_\theta^G$ is found as
		\begin{align}
			\Omega_{\theta}^{G}&=\frac{1}{4 \pi^2 \sigma_x^2 \sigma_y^2} \iint \rho_1(x-\Delta_x,y-\Delta_y,\theta)\rho_2(x,y,\theta) \, dx \, dy\nonumber\\
			&=\Omega_{\theta,x}(\Delta_x,\theta)\, \Omega_{\theta,y}(\Delta_y,\theta),
			\label{eq:25}
		\end{align}
		where $\Omega_{\theta,x}^G$ and $\Omega_{\theta,y}^G$ are the overlap integral in the horizontal and vertical direction, respectively, and defined as:
			\begin{equation}
			\begin{split}
			\Omega_{\theta,x}(\Delta_x,\theta)=\frac{1}{\sqrt{2\pi}\sqrt{(\sigma_x^2+\sigma_y^2)+(\sigma_x^2-\sigma_y^2)\cos(\theta)}}\\ \times\exp \left(\frac{-\Delta_x^2}{2\left((\sigma_x^2+\sigma_y^2)+(\sigma_x^2-\sigma_y^2)\cos(\theta)\right)}\right);\\
			\Omega_{\theta,y}(\Delta_y,\theta)=\frac{1}{\sqrt{2\pi}\sqrt{(\sigma_x^2+\sigma_y^2)-(\sigma_x^2-\sigma_y^2)\cos(\theta)}}\\ \times\exp \left(\frac{-\Delta_y^2}{2\left((\sigma_x^2+\sigma_y^2)-(\sigma_x^2-\sigma_y^2)\cos(\theta)\right)}\right).
			\end{split}
			\label{eq:26}
			\end{equation}
		 
		 The tilt angle leads to non-factorizable transverse densities in the lab frame $XYZ$, but their resultant overlap integral is factorizable in terms of beam separation $\Delta_{x}$ and $\Delta_{y}$ with Gaussian form as in equations \eqref{eq:25} and \eqref{eq:26}. Thus, the horizontal and vertical convolved beam sizes $\Sigma_{\theta,x}^G$ and $\Sigma_{\theta,y}^G$ are not coupled, and they can be determined via two separate one-dimensional vdM scans as
		\begin{equation}
		 	\begin{split}
		 	\Sigma^G_{\theta,x}(\theta)&=\frac{1}{\sqrt{2\pi}}
		 		 \frac{\int\Omega_{\theta,x}^G(\Delta_x,\theta)\ d\Delta_x}{\Omega_{\theta,x}^G(0,\theta)}\\
	  		&=\sqrt{(\sigma_x^2+\sigma_y^2)+(\sigma_x^2-\sigma_y^2)\cos(\theta)}\ ;\\
		  	\Sigma^G_{\theta,y}(\theta)&=\frac{1}{\sqrt{2\pi}} \frac{\int \Omega_{\theta,y}^G(\Delta_y,\theta\ )\ d\Delta_y}{\Omega_{\theta,y}^G(0,\theta)}\\
		 	&=\sqrt{(\sigma_x^2+\sigma_y^2)-(\sigma_x^2-\sigma_y^2)\cos(\theta)}\ .
		 	\end{split}
		 	\label{eq:27}
		 	\end{equation}
		 
\begin{figure}
		 	\centering
		 	\includegraphics[width=0.9\linewidth]{{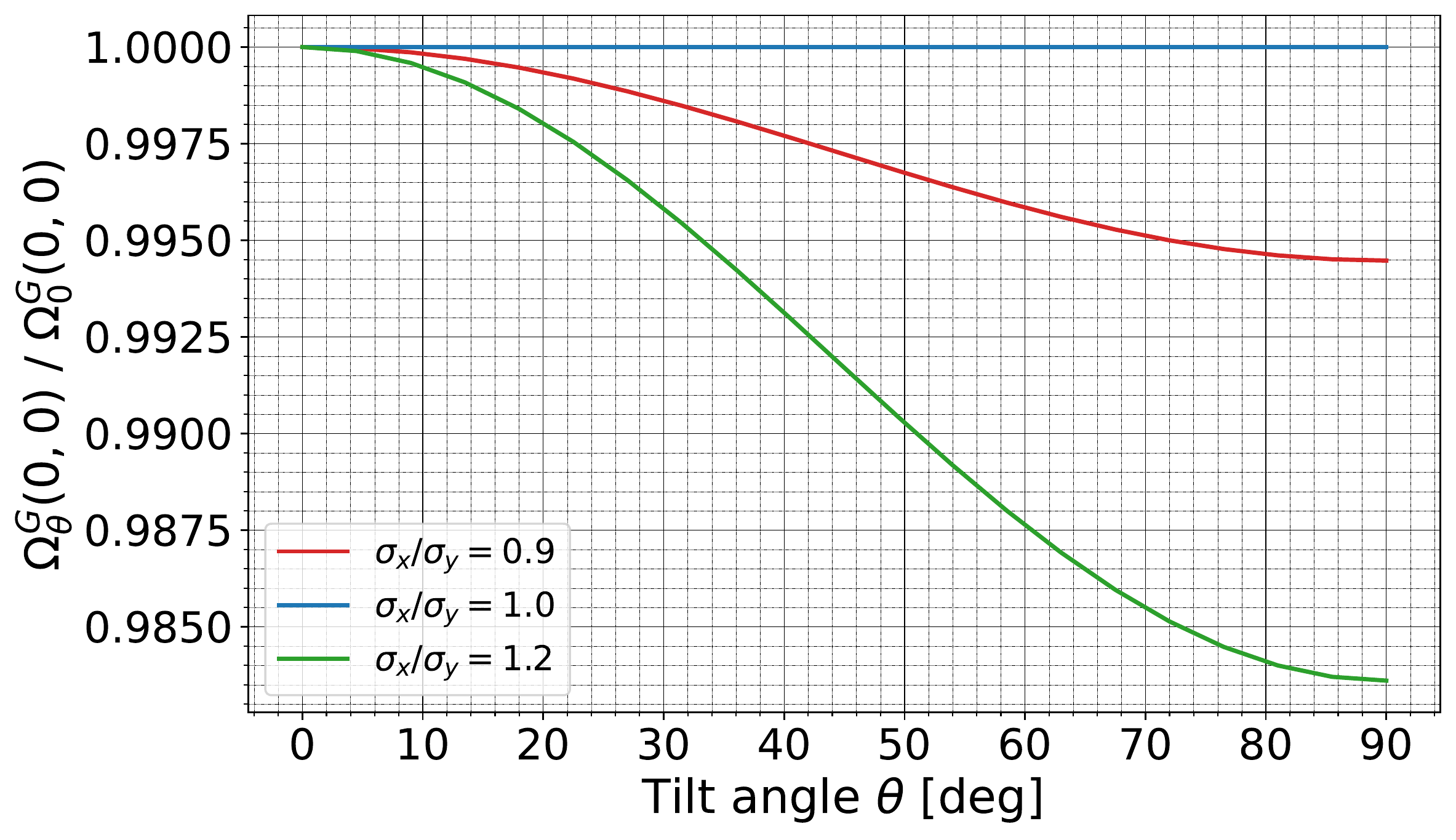}}
		 	\caption{The ratio of maximum overlap integral at a tilt angle to that at zero tilt angle of Gaussian bunches $\frac{\Omega_\theta^G}{\Omega_0^G}$ at tilt angle $\theta$ in the range $0$ to ${90}^o$ for different horizontal to vertical bunch dimension ratios $\frac{\sigma_x}{\sigma_y}= 0.9$, $1$ and $1.2$  for $\sigma_y=100$ $\mu m$ at zero separation}
		 	\label{fig:6}
		 \end{figure}

		 In principle, the tilt angle $\theta$ leads to a reduction in the overlap integral, which can be considered a geometrical reduction factor due to the $x-y$ coupling in the transverse particle densities. Figure \ref{fig:6} shows the maximum overlap integral $\Omega_\theta^G$ at a tilt angle $\theta$ normalized to that at zero tilt angle $\Omega_0^G$ for Gaussian bunches with different horizontal to vertical bunch dimension ratios with tilt angle $\theta$ in the range $0$ to ${90}^o$. For round bunches, $\frac{\sigma_x}{\sigma_y}=1$, the tilt angle does not produce any effect. As the $\frac{\sigma_x}{\sigma_y}$ ratio increases, the reduction effect increases.
	 
		 \subsubsection{Non-Gaussian tails}
\begin{figure}
			\begin{center}
				\subfloat[\label{subfig:7a}]{\includegraphics[width=0.9\linewidth]{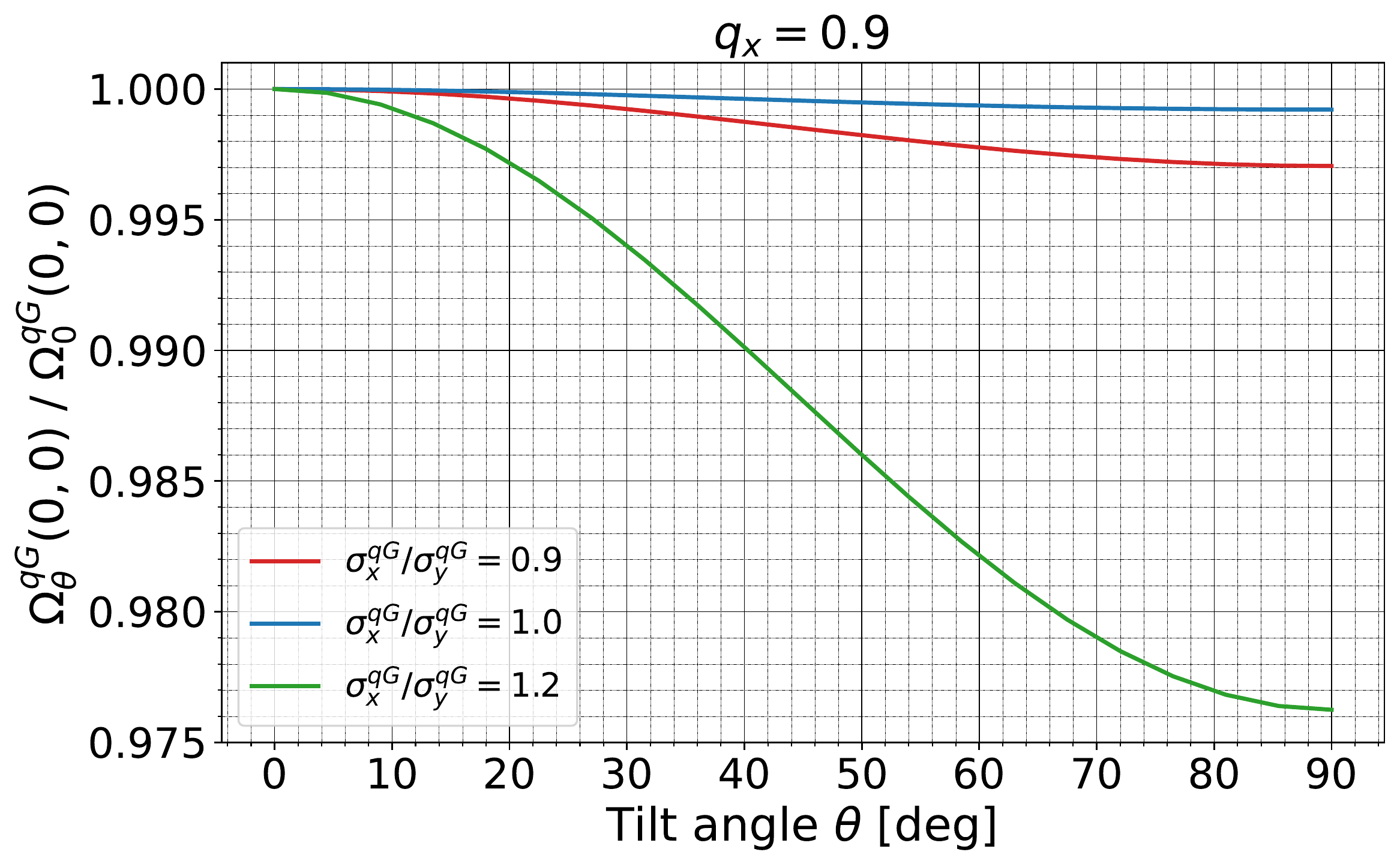}}\	\subfloat[\label{subfig:7b}]{\includegraphics[width=0.9\linewidth]{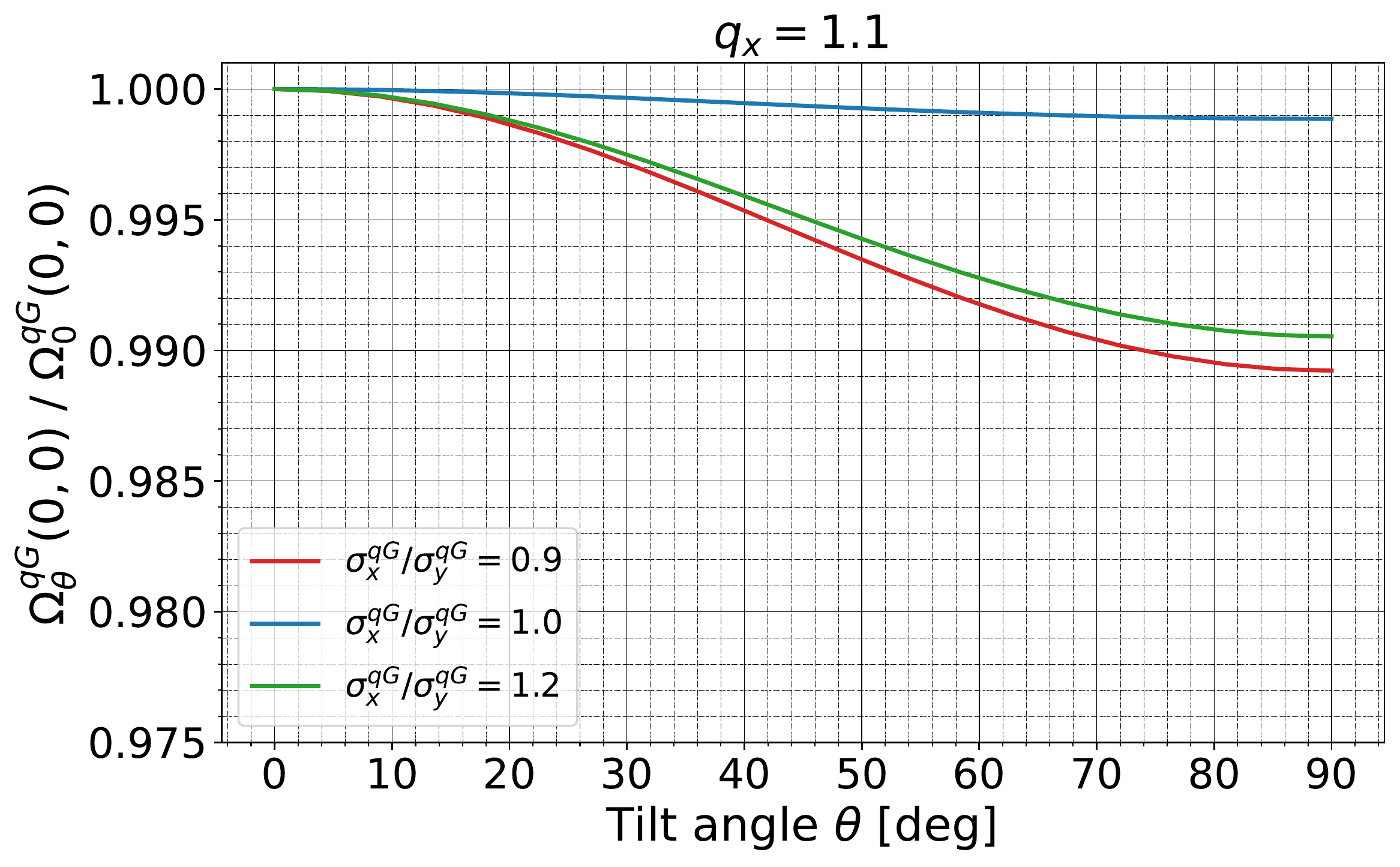}}
			\end{center}
			\caption{The ratio of maximum overlap integral of q-Gaussian bunches at a tilt angle $\theta$ to that at zero tilt angle $\frac{\Omega_\theta^{qG}}{\Omega_0^{qG}}$ with horizontal tail density $q_x=0.9$ (\ref{subfig:7a}) and $q_x=1.1$ (\ref{subfig:7b}) at tilt angle $\theta$ in the range $0$ to ${90}^o$ for different horizontal to vertical bunch dimension ratios $\frac{\sigma_x^{qG}}{\sigma_y^{qG}}=0.9$, $1$ and $1.2$  , where the vertical bunch dimension $\sigma_y^{qG}=100$ $\mu m$}
			\label{fig:7}
		\end{figure}

		 Since the bunches with non-Gaussian tails can be represented by q-Gaussian distribution as mentioned earlier, let’s assume two q-Gaussian bunches $\rho_{1,2}$ with equal bunch dimensions and  tail densities in the horizontal and vertical directions such that ($\sigma_{1x}^{qG}=\sigma_{2x}^{qG}=\sigma_x^{qG}$; $\sigma_{1y}^{qG}=\sigma_{2y}^{qG}=\sigma_y^{qG}$; $q_{1x}=q_{2x}=q_x;$ and $q_{1y}=q_{2y}=q_y$) collide at zero crossing angle, using similar frames of reference and coordinates as in equation \eqref{eq:24}, the particle densities in the lab frame are defined as:
		     \begin{equation}
		 	\begin{split}
		 		\rho_1(x,y,&\theta)=QG\left(x\cos \left( \frac{\theta}{2} \right) + y\sin \left(\frac{\theta}{2}\right); q_x,\beta^{qG}_x\right) \\
		 		&\times QG\left( - x\sin \left( \frac{\theta}{2} \right) + y\cos \left(\frac{\theta}{2}\right);q_y,\beta_y^{qG}\right);\\
		 		\rho_2(x,y,&\theta)=QG\left(x\cos \left( \frac{\theta}{2} \right) - y\sin \left(\frac{\theta}{2}\right); q_x,\beta_x^{qG}\right) \\
		 		&\times QG\left( - x\sin \left( \frac{\theta}{2} \right) - y\cos \left(\frac{\theta}{2}\right);q_y,\beta_y^{qG}\right),
		 	\end{split}
		 	\label{eq:28}
		 \end{equation}
		and the overlap integral $\Omega_\theta^{qG}$ is given by
		 \begin{align}
		 	&\Omega_\theta^{qG}(\Delta_x,\Delta_y,\theta)=\frac{\beta^{qG}_x\beta^{qG}_y}{C_x^{{qG}^2}C_y^{{qG}^2}}\nonumber\\
			&\times \mathop{\mathlarger{\mathlarger{\iint}}}
			e_{q_x}\left(\beta^{qG}_x\left( x\cos\left(\frac{\theta}{2}\right)-y\sin\left(\frac{\theta}{2}\right)\right)^2\right)\nonumber\\
			&\times e_{q_x}\left(\beta^{qG}_x\left( (x-\Delta_x)\cos\left(\frac{\theta}{2}\right)+(y-\Delta_y)\sin\left(\frac{\theta}{2}\right)\right)^2\right)\nonumber\\
			&\times e_{q_y}\left(\beta^{qG}_y\left( (x-\Delta_x)\sin\left(\frac{\theta}{2}\right)+(y-\Delta_y)\cos\left(\frac{\theta}{2}\right)\right)^2\right)\nonumber\\
			&\times e_{q_y}\left(\beta^{qG}_y\left( x\sin\left(\frac{\theta}{2}\right)-y \cos\left(\frac{\theta}{2}\right)\right)^2\right) \, dx \, dy. 
			\label{eq:29}
		\end{align}
		The solution of equation \eqref{eq:29} is hard to find analytically; therefore, numerical integration techniques are used to estimate the impact of tilt angle on the overlap integral of q-Gaussian bunches. Similar to Gaussian, the existence of the tilt angle leads to a reduction in the overlap integral. The dependence of this reduction on the tilt angle at different horizontal tail densities $q_x$ and bunch dimension ratios $\frac{\sigma_x^{qG}}{\sigma_y^{qG}}$ is shown in Fig. \ref{fig:7}, where the vertical tail density is assumed to be Gaussian, $q_y=1$. Figure \ref{fig:7} shows that for the bunch dimension ratio $\frac{\sigma_x}{\sigma_y}=1$, the reduction effect is exist for both light- and heavy-tailed bunches, where it is larger for heavy-tailed bunches. For bunches with heavy-tailed horizontal density $q_x>1$, if  $\frac{\sigma_x}{\sigma_y}>1$, the reduction effect is less than Gaussian while for $\frac{\sigma_x}{\sigma_y}<1$, the reduction effect is higher than Guassian. For bunches with light-tailed horizontal density $q_x<1$, if  $\frac{\sigma_x}{\sigma_y}>1$, the reduction effect is higher than Gaussian while for $\frac{\sigma_x}{\sigma_y}<1$, the reduction effect is less than Guassian.
		It worth nothing that for q-Gaussian bunches, the tilt angle leads to a non-factorable overlap integral $\Omega_\theta^{qG}$ in terms of beam separation $\Delta_{x}$ and $\Delta_{y}$, which leads to a non-factorization bias in vdM scan. The detailed study of this effect is abroad of the topic and will be published elsewhere.
	
\section{Application}
\label{sec:4}
The previous section estimates how the non-Gaussian tails affect the overlap integral and convolved beam size, where the q-Gaussian bunches are used as a more realistic approximation to the actual bunches. In this section, the influence of non-Gaussian tails on quantities derived from van-der-Meer (vdM) scan is investigated. Namely, a ``{\em toy}''  vdM scan is modelled by q-Gaussian bunches, and the resultant scan data is fitted by Gaussian, double Gaussian and q-Gaussian fit models to estimate the precision of the Gaussian-based models when it is applied to bunches with non-Gaussian tails. After that, the procedure is tested using beam overlap width measurements performed in 2015 CMS vdM scan program.

	\subsection{Toy van-der-Meer scan}
	\label{subsec:4.1}
	The ``{\em toy}” van-der-Meer scan was performed by calculating the overlap integral of two q-Gaussian bunches with equal bunch dimension $\sigma_u^{qG}$  and tail density $q$, where the overlap integral is calculated by the analytical formulas of $\Omega_u^{qG}$, equations \eqref{eq:17} and \eqref{eq:19}, for different separations $\Delta_u$ from  $0$ to $6$ $\sigma_u^{qG}$, where 60 points of $\left(\Delta_u, \Omega_u^{qG}(\Delta_u;q)\right)$ is obtained, and the convolved beam size of these bunches $\Sigma_u^{qG}$ is calculated by the analytical formula \eqref{eq:27}.
	For the vdM scan, the standard fit model for application is Gaussian $f^{G}$, equation \eqref{eq:30}, but since the simple Gaussian does not adequately fit the scan data, especially the tails; therefore, a double Gaussian fit  $f^{DG}$, equation \eqref{eq:31},  with two different widths is widely used in RHIC \cite{r11} and  LHC \cite{r14}, where the Gaussian with smaller width $\sigma_1$ fits the core, and the Gaussian with the larger width $\sigma_{2}$ fits the tails, and $\epsilon$ is the fraction of Gaussian with the smaller width. The convolved beam size $\Sigma_u^{DG}$ $(u=x,y)$ is defined as in equation \eqref{eq:32}.
	\begin{equation}
		f^G(\Delta_u)=\frac{1}{\sqrt{2\pi}\Sigma_u^G}\exp\left( -\frac{\Delta_u^2}{2\Sigma_u^{G^2}}\right).
		\label{eq:30}
	\end{equation}
	
	\begin{align}
			f^{DG} (\Delta_u )=\frac{1}{\sqrt{2\pi}} \bigg[\frac{\epsilon}{\sigma_{1u}}&\exp\left( -\frac{\Delta_u^2}{2\sigma_{1u}^{2}}\right)\nonumber\\			    &+\frac{1-\epsilon}{\sigma_{2u}}exp\left( -\frac{\Delta_u^2}{2\sigma_{2u}^{2}}\right) \bigg].
	\label{eq:31}
	\end{align}
	
	\begin{equation}
		\Sigma_u^{DG}=\frac{\sigma_{1u}\sigma_{2u}}{\epsilon_u\sigma_{2u}+(1-\epsilon_u)\sigma_{1u}}.
	\label{eq:32}
	\end{equation}

	Based on the q-Gaussian bunches assumption and the derived analytical formulae of the overlap integral $\Omega_u^{qG}$, a vdM scan fit model is proposed to account for the tail populations. The model is based on the ability of the q-Gaussian distribution to describe various tails, ranging from finite light tails for $q<1$ to heavy tails for $q>1$. The proposed model $f_u^{qG}$ is defined in terms of $q$ and $\Sigma_u^{qG}$  as:
	    \begin{equation}
	f_u^{qG}(\Delta_u;q)=\frac{1}{C_u^{qG}\sqrt{5-3q}\Sigma_u^{qG}}e_q\left( - \frac{\Delta_u^2}{(5-3q)\Sigma_u^{qG^2}}\right),
		\label{eq:33}
		\end{equation}
	where $e_q$ and $C^{qG}$  are defined by equations \eqref{eq:9} and \eqref{eq:10}.
	
	The toy scan is conducted for light-tailed bunches with ``$q=0.8$'' and heavy-tailed bunches with ``$q=1.2$'', with equal bunch dimensions $\sigma_u^{qG}=100$ $\mu m$ , The resultant toy scans data are then fitted by models \eqref{eq:30}, \eqref{eq:31}, and \eqref{eq:33} using least-squares minimization by the trust region reflective method from Non-Linear Least-Squares Minimization and Curve-Fitting Python Package ``{\em lmfit}''. Since the overlap integral of light-tailed bunches has underpopulated tails, the double Gaussian model \eqref{eq:31} was applied only for heavy-tailed bunches, as the concept of its application does not coincide with the light-tailed bunches.
	
\begin{figure}
			\subfloat[\label{subfig:8a}]{\includegraphics[width=0.9\linewidth]{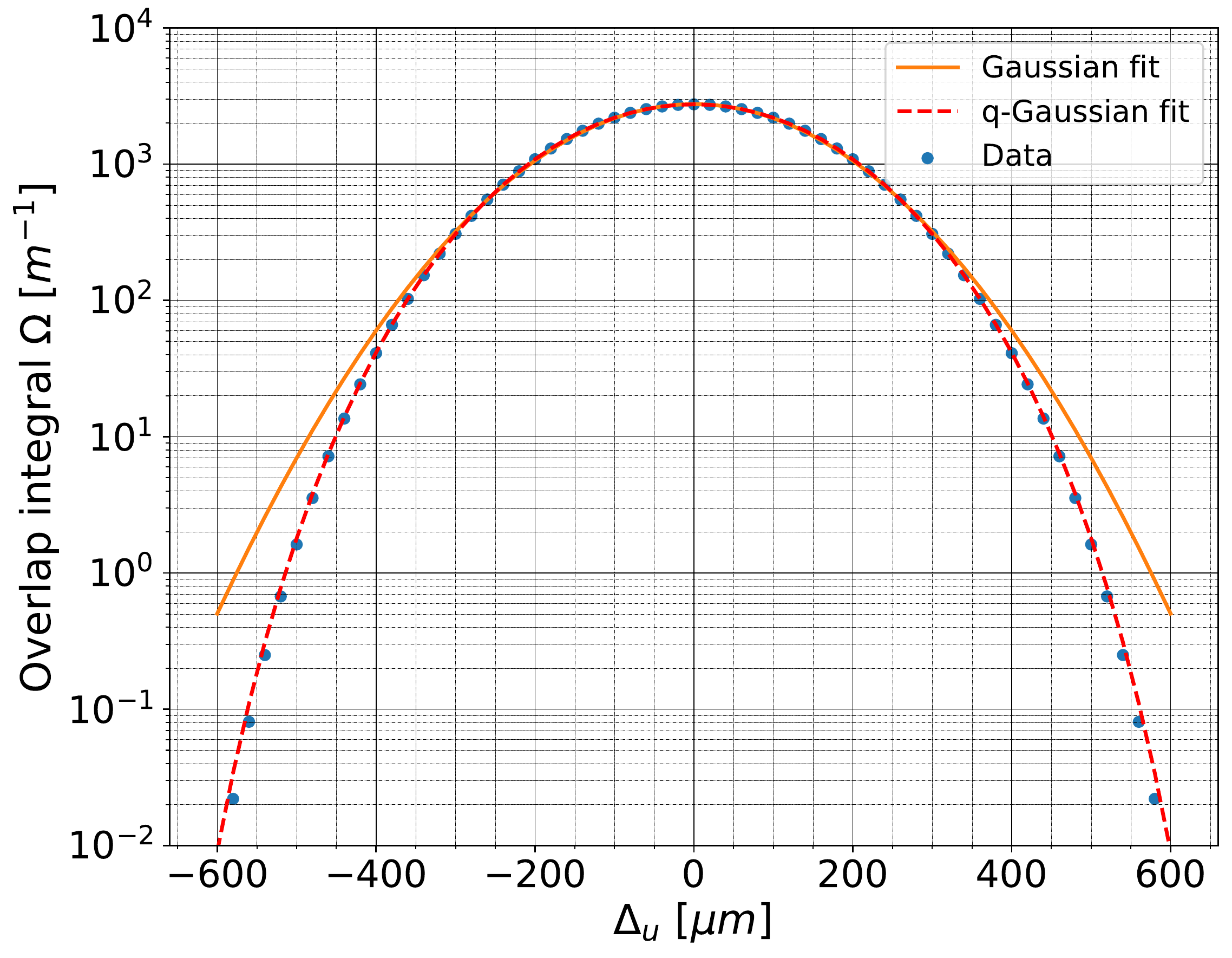}}\\	\subfloat[\label{subfig:8b}]{\includegraphics[width=0.9\linewidth]{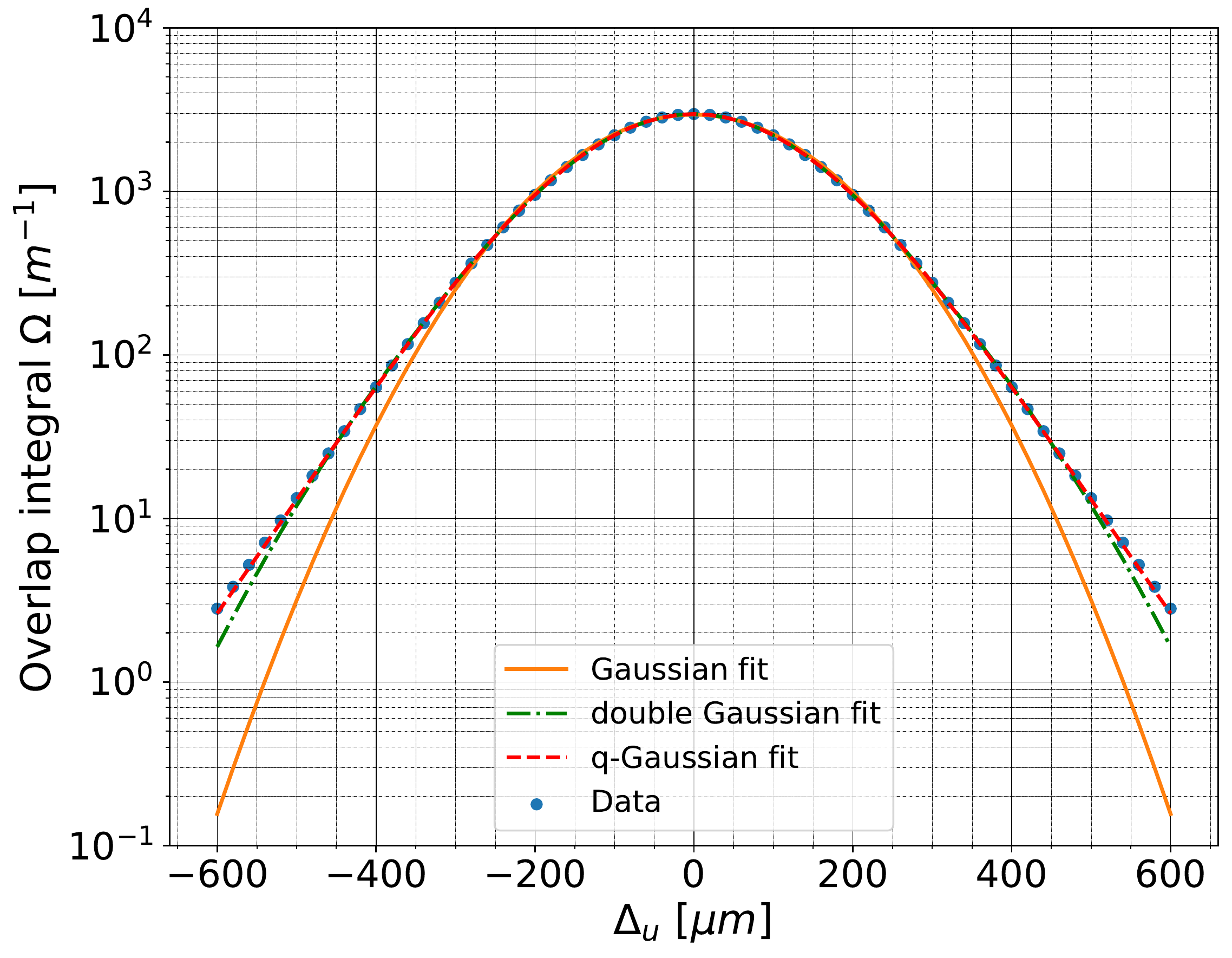}}
		\caption{The fitting of the overlap integral of two q-Gaussian bunches with equal bunch size $\sigma^{qG}_{u}=100\ \mu m$ and tail density $q$ obtained at separation $\Delta_u$ in the range from $0$  to $6$ $\sigma_u^{qG}$ during the ``{\em toy}'' vdM scan. Two cases were considered: light-tailed bunches ``$q=0.8$'' (\ref{subfig:8a}) and heavy-tailed bunches ``$q=1.2$'' (\ref{subfig:8b}), where the following fit models were applied: Gaussian \eqref{eq:30}; double Gaussian \eqref{eq:31}; and q-Gaussian \eqref{eq:33}}
		\label{fig:8}
	\end{figure}

	The fitting of vdM toy scan data is shown in Fig. \ref{fig:8}. For light-tailed beams ``$q=0.8$'', Fig. \ref{subfig:8a}, the Gaussian model failed to fit the data at small and large separations, where it overestimates the data core and tails; on the other hand, the q-Gaussian model fits the data well. For heavy-tailed beams at ``$q=1.2$'', Fig. \ref{subfig:8b}, the Gaussian model underestimates the core and the tails of the data; on the other hand, the double Gaussian and q-Gaussian models provide a good description of the data. Compared to Gaussian and double Gaussian fits, the q-Gaussian fit provides the best description of the data, especially for the tails. The fitting parameters are summarized in Table \ref{tab:2}. The goodness of fit statistics are based on the root mean square error (RMSE) and $R^2$  statistics and the deviation of the predicted values of $\Sigma_u^{qG}$ and the maximum $\Omega_u^{qG}$ (at zero separation) are summarized in Table  \ref{tab:3}. 
	\begin{table*}
		\caption{The estimated fitting parameters for the ``{\em toy}'' vdM scan data of two q-Gaussian bunches with equal size $\sigma_{u}^{qG}=100\ \mu m$ and tail density  $q$. Two cases were considered: light-tailed beams ``$q=1.2$'' and heavy-tailed beams ``$q=1.2$'', and the following fit models were applied: Gaussian \eqref{eq:30}; double Gaussian \eqref{eq:31}; and q-Gaussian \eqref{eq:33}}
		\label{tab:2}
		\begin{tabular*}{\textwidth}{@{\extracolsep{\fill}}ll@{}}
			\begin{tabular*}{\columnwidth}{@{\extracolsep{\fill}}lccc@{}}
				\hline
				\multicolumn{4}{@{}c}{Light-tailed $q=0.8$} \\
				\hline
				\multicolumn{3}{@{}c}{Fit parameters} & SD errors [${10}^{-4}$] \\
				\hline
				Gaussian             	         & $\Sigma_u^G\, [\mu m]$    & $144.59$ & $0.28037$ \\
				Double  Gaussian				 & $\epsilon$ 				 & ---  	& ---\\
											   	 & $\sigma_{1u}^G\, [\mu m]$ & ---		& ---\\
												 & $\sigma_{2u}^G\, [\mu m]$ & ---		& ---\\
				q-Gaussian						 & $q$						 & $0.9226$	& $0.00021$ \\
				  								 & $\Sigma_u^{qG}\, [\mu m]$ & $141.48$	& $0.00929$ \\
				\hline
			\end{tabular*} &
			\begin{tabular*}{\columnwidth}{@{\extracolsep{\fill}}lccc@{}}
				\hline
				\multicolumn{4}{@{}c}{Heavy-tailed $q=1.2$} \\
				\hline
				\multicolumn{3}{@{}c}{Fit parameters} & SD errors [${10}^{-4}$] \\
				\hline
				Gaussian             	         & $\Sigma_u^G\, [\mu m]$    & $137.4$  & $0.44499$ \\
				Double  Gaussian				& $\epsilon$ 				 & $0.206$  & $0.00730$ \\
												& $\sigma_{1u}^G\, [\mu m]$ & $115.6$	& $1.17431$ \\
												& $\sigma_{2u}^G\, [\mu m]$ & $167.7$	& $1.27512$ \\
				q-Gaussian						 & $q$						 & $1.1257$	& $0.00017$ \\
												& $\Sigma_u^{qG}\, [\mu m]$ & $141.24$	& $0.01224$ \\
				\hline
			\end{tabular*}
		\end{tabular*}
	\end{table*}
	\begin{table*}
		\caption{Goodness of fit statistics based on RMSE and $R^2$, and the deviation of $\Sigma_u^{fit\ model}$  and $\Omega_u^{fit\ model}$, determined from the fitting of the ``{\em toy}'' vdM scan data, form their analytical values  $\Sigma_u^{Analytical}$, equation \eqref{eq:22}, and the maximum $\Omega_u^{Analytical}$ (at zero separation), equations \eqref{eq:17} and \eqref{eq:19}, for two cases: light-tailed bunches ``$q=0.8$'' and heavy-tailed bunches ``$q=1.2$''}
		\label{tab:3}
		\begin{tabular*}{\textwidth}{@{\extracolsep{\fill}}lccccc@{}}
			\hline
			Fitting model & RMSE & $R^2$ & Adj. $R^2$ & $\frac{\Sigma_u^{fit\ model}}{\Sigma_u^{Analytical}}-1\, [\%]$ & $\frac{\Omega_u^{fit\ model}}{\Omega_u^{Analytical}}-1\, [\%]$ \\
			\hline
			\multicolumn{6}{@{}c}{Light-tailed $q=0.8$}\\
			\hline
			Gaussian 	&	$2.65232$ &	$0.9997128$ &	$0.9997128$ &	$5.784$ &	$0.461$ \\
			q-Gaussian  &	$0.33639$ & $0.9999999$ &	$0.9999999$ &	$3.503$ &	$-0.022$ \\
			\hline
			\multicolumn{6}{@{}c}{Heavy-tailed $q=1.2$}\\
			\hline
			Gaussian    	&	$29.031$ &	$0.9991700$ &	$0.9991700$ &	$-8.625$ &	$-0.771$ \\
			Double Gaussian &	$1.42527$ &	$0.9999980$ &	$0.9999979$ &	$-9.295$ &	$-0.037$ \\
			q-Gaussian      &	$0.31541$ & $0.9999999$ &	$0.9999999$ &	$-4.197$ &	$0.018$ \\
			\hline
		\end{tabular*}
	\end{table*}
	
	Based on the previous results, the q-Gaussian fit model ($f^{qG}$), equation \eqref{eq:33}, provides the best fit for the scan data and predicts the convolved beam size and the overlap integral with high precision with deviations less than $\pm5\%$ and $\pm0.025\%$, respectively. Double Gaussian fit model ($f^{DG}$), equation \eqref{eq:31}, provides a good fit for the data, and predicts overlap integral with a deviation less than $0.04\%$, but it can only be applied for infinite heavy-tailed bunches, since for bounded light-tailed bunches, it gives the same results as Gaussian, and it fails to provide any better predictions than single Gaussian. The high deviation of the Gaussian-based models compared to the q-Gaussian model is because the Gaussian-based models do not account for the relation between the convolved beam size and the tail density, even for  $f^{DG}$,  which was only successful for the heavy-tailed bunches.
	 
\begin{figure}
			\subfloat[\label{subfig:9a}]{\includegraphics[width=0.9\linewidth]{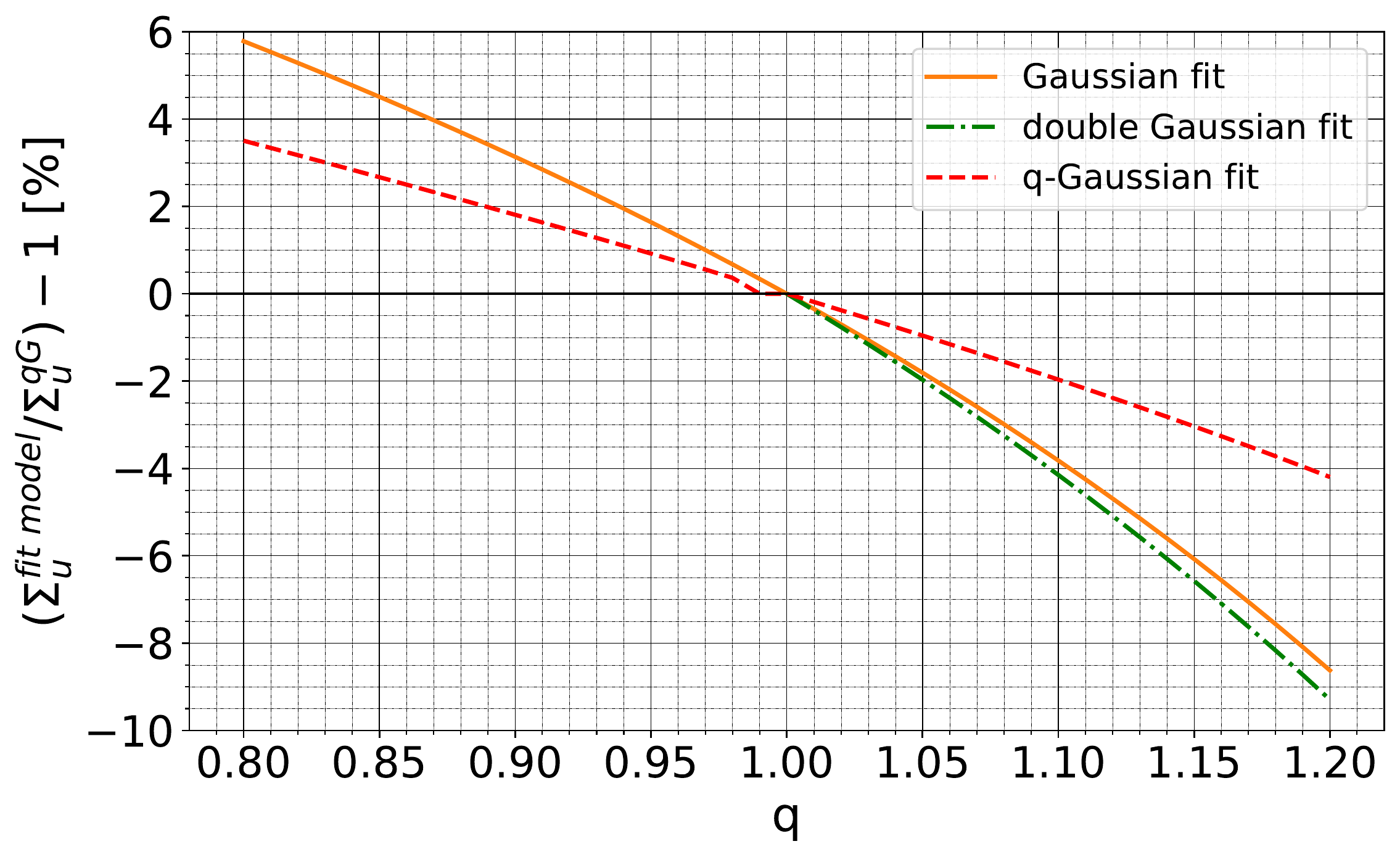}}\\	\subfloat[\label{subfig:9b}]{\includegraphics[width=0.9\linewidth]{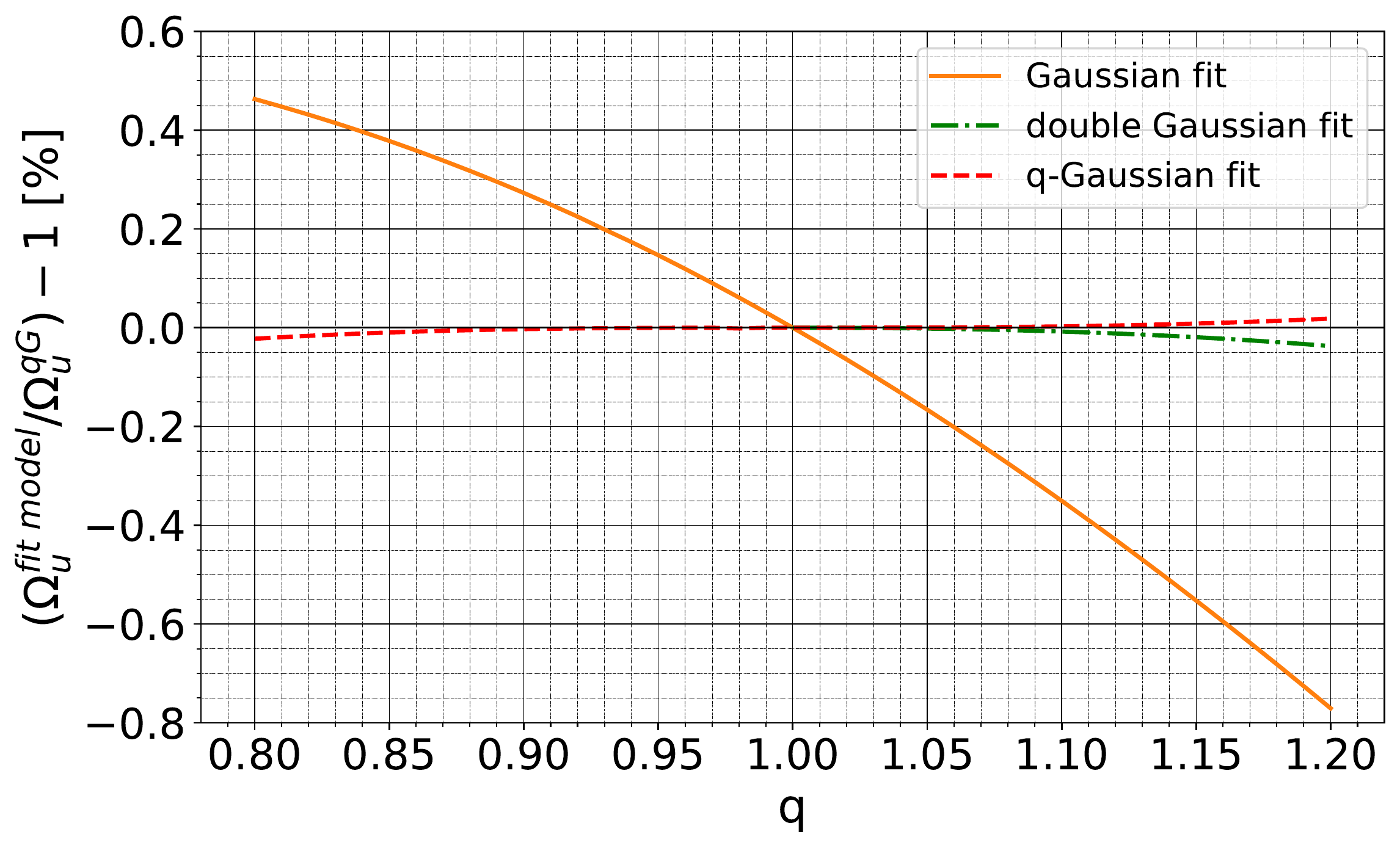}}
		\caption{The deviation of the convolved beam size $\Sigma_u^{fit\ model}$ (\ref{subfig:9a}) and the overlap integral $\Omega_u^{fit\ model}$ (\ref{subfig:9b}), determined by the different fit models, from their respective analytical values $\Sigma_u^{Analytical}$ and $\Omega_u^{Analytical}$, calculated by equation \eqref{eq:20} and equations \eqref{eq:14}, \eqref{eq:17} and \eqref{eq:19}, for ``{\em toy}'' vdM scan of two equal-size q-Gaussian bunches with tail density $q$ in the range $0.8$ to $1.2$}
		\label{fig:9}
	\end{figure}

\begin{figure}
			\subfloat[\label{subfig:10a}]{\includegraphics[width=0.9\linewidth]{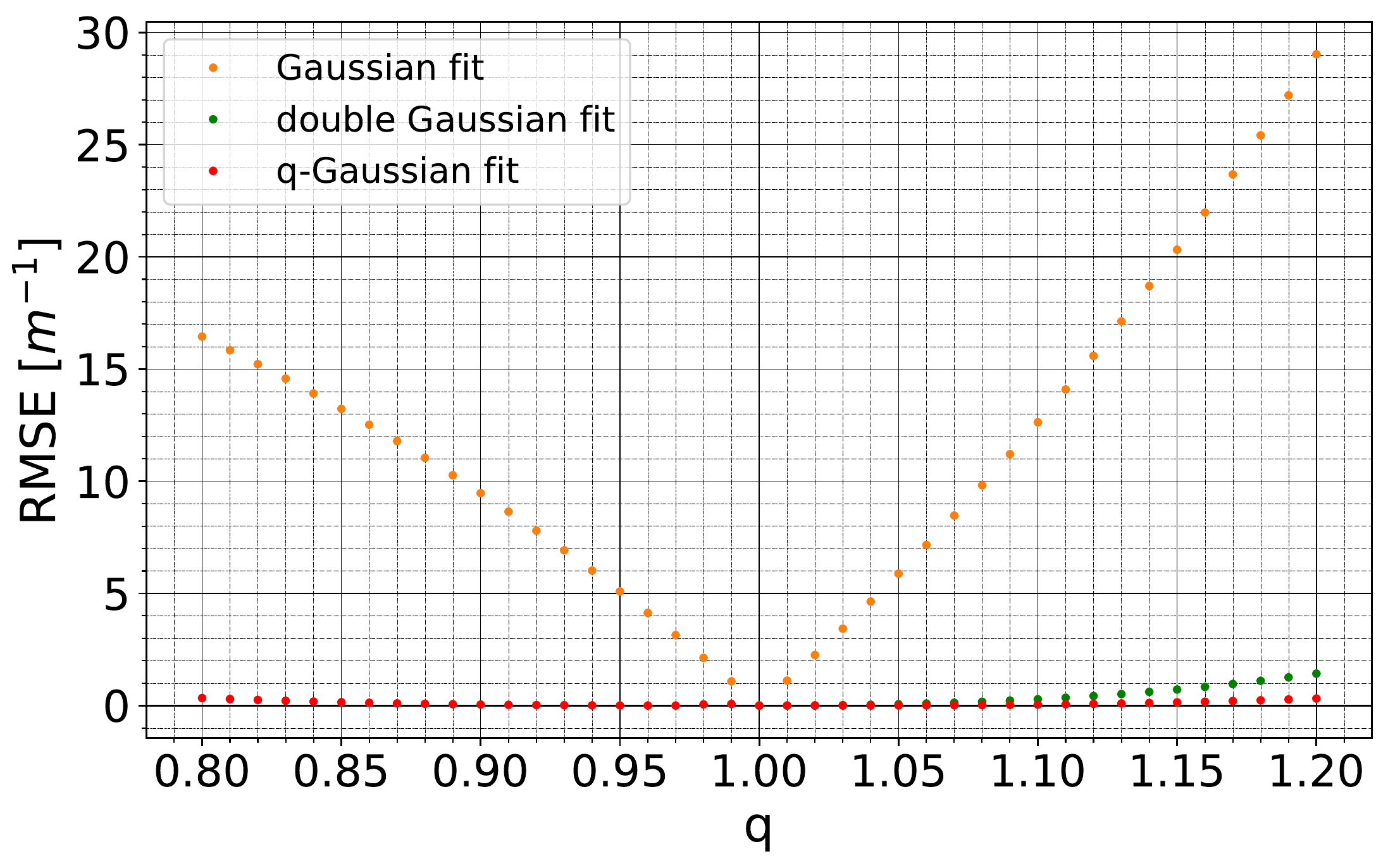}} \\	\subfloat[\label{subfig:10b}]{\includegraphics[width=0.9\linewidth]{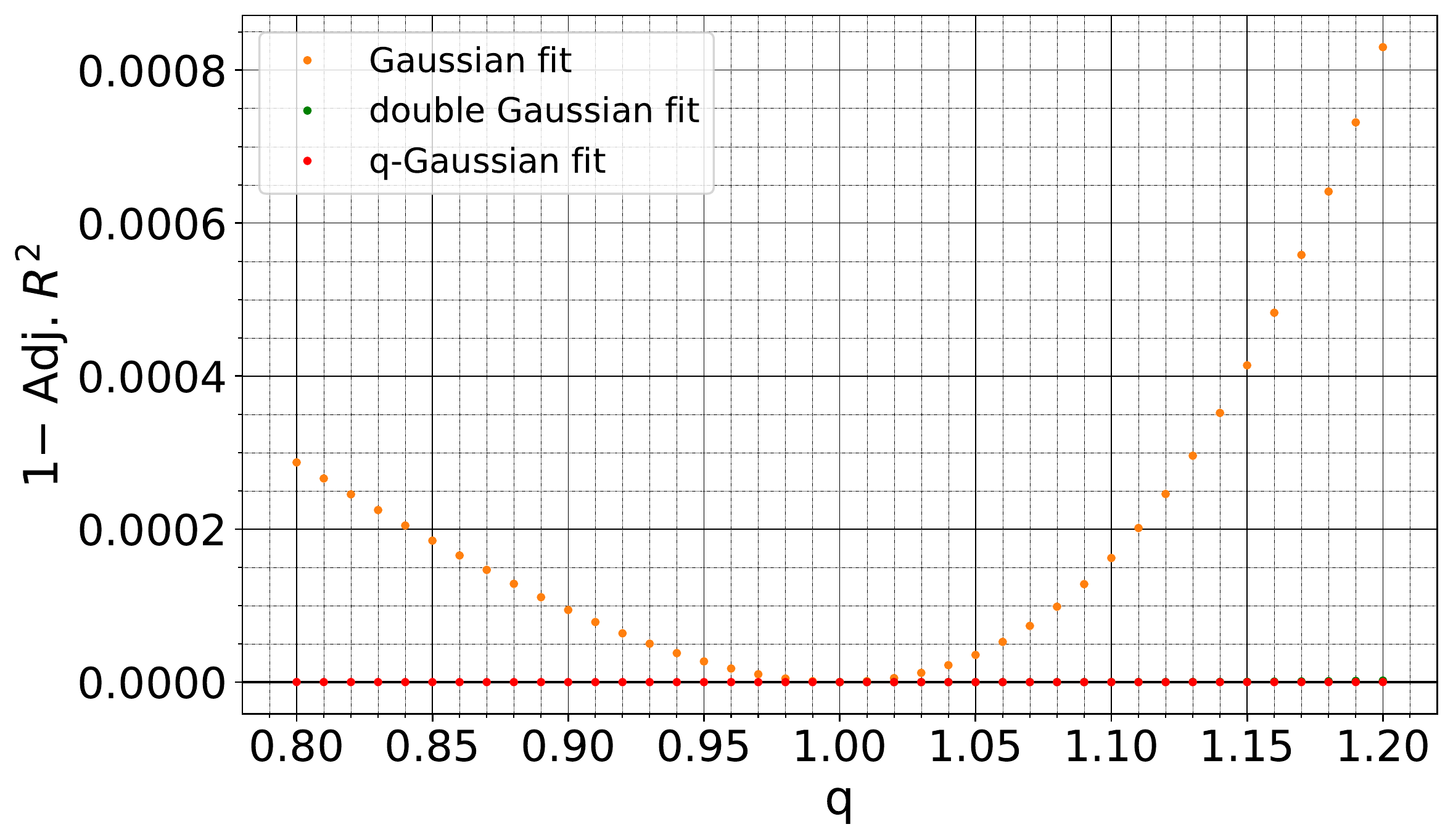}}
		\caption{Goodness of fit analysis based on RMSE (\ref{subfig:10a}) and Adj. $R^2$ (\ref{subfig:10b}) for different fit models of toy vdM scan of two equal-size q-Gaussian bunches with tail density $q$  in the range 0.8  to 1.2}
		\label{fig:10}
		\end{figure}

	To investigate the effect of the non-Gaussian tails on the accuracy of prediction of different fit models, the toy vdM scan was performed with tail density $q$ in the range $0.8$ to $1.2$. Figure \ref{fig:9} show the deviation of the predicted convolved beam size  $\Sigma_u^{fit\ model}$  and overlap integral  $\Omega_u^{fit\ model}$  obtained by the previous fit models from their respective analytical values calculated by equations \eqref{eq:17} and \eqref{eq:19}, respectively. For light-tailed bunches $f^{G}$  and  $f^{DG}$ overestimate the convolved beam size and the overlap integral, while for heavy-tailed beams $f^{G}$, $f^{DG}$  and $f^{qG}$  underestimate them. When $q$  tends to 1, all models have predictions close to the analytical values. The goodness of fit statistics based on RMSE and Adj. $R^2$  at various tail densities $q$ is shown in Fig. \ref{fig:10}.  
	
	Based on Figs. \ref{fig:9} and \ref{fig:10}, the q-Gaussian fit model, equation \eqref{eq:33}, can predict the convolved beam size and the overlap integral with higher accuracy than Gaussian and double Gaussian models; it also provides a better description of the scan data according to the goodness of fit statistics. It can be applied for both light- and heavy-tailed bunches. Compared to the analytical solution in equations \eqref{eq:17} and \eqref{eq:19}, the q-Gaussian fit model represents a good and straightforward approximation.
	
	\subsection{Influence of the non-Gaussian tails at the vdM scan at LHC}
	\label{subsec:4.2}
	
	The beam overlap $\Omega$ is not a measurable quantity. The real measurable is response rate $R$ mentioned above. The response $R$ is proportional to $\Omega$ because $R$ characterizes the flux of collision product whereas $\Omega$ characterizes the intensity of collisions. Therefore, the measured rates $R$ in the vdM scan form scan shape that is similar to the scan shape in terms of the corresponding $\Omega$ considered above. In this subsection, we apply models developed before to response rates $R$.
	
\begin{figure}
		\subfloat[\label{subfig:11a}]{\includegraphics[width=0.9\linewidth]{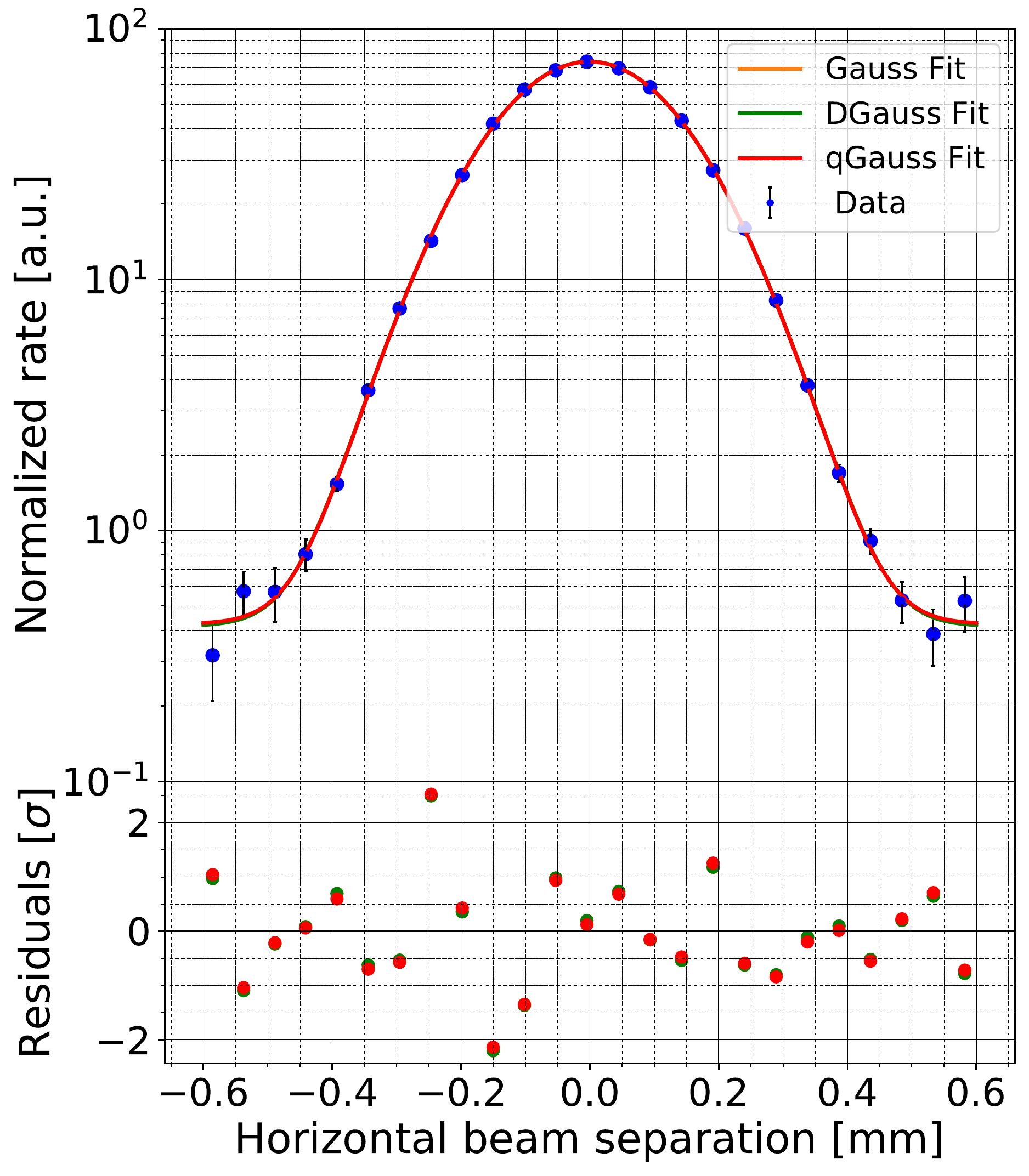}}\\	\subfloat[\label{subfig:11b}]{\includegraphics[width=0.9\linewidth]{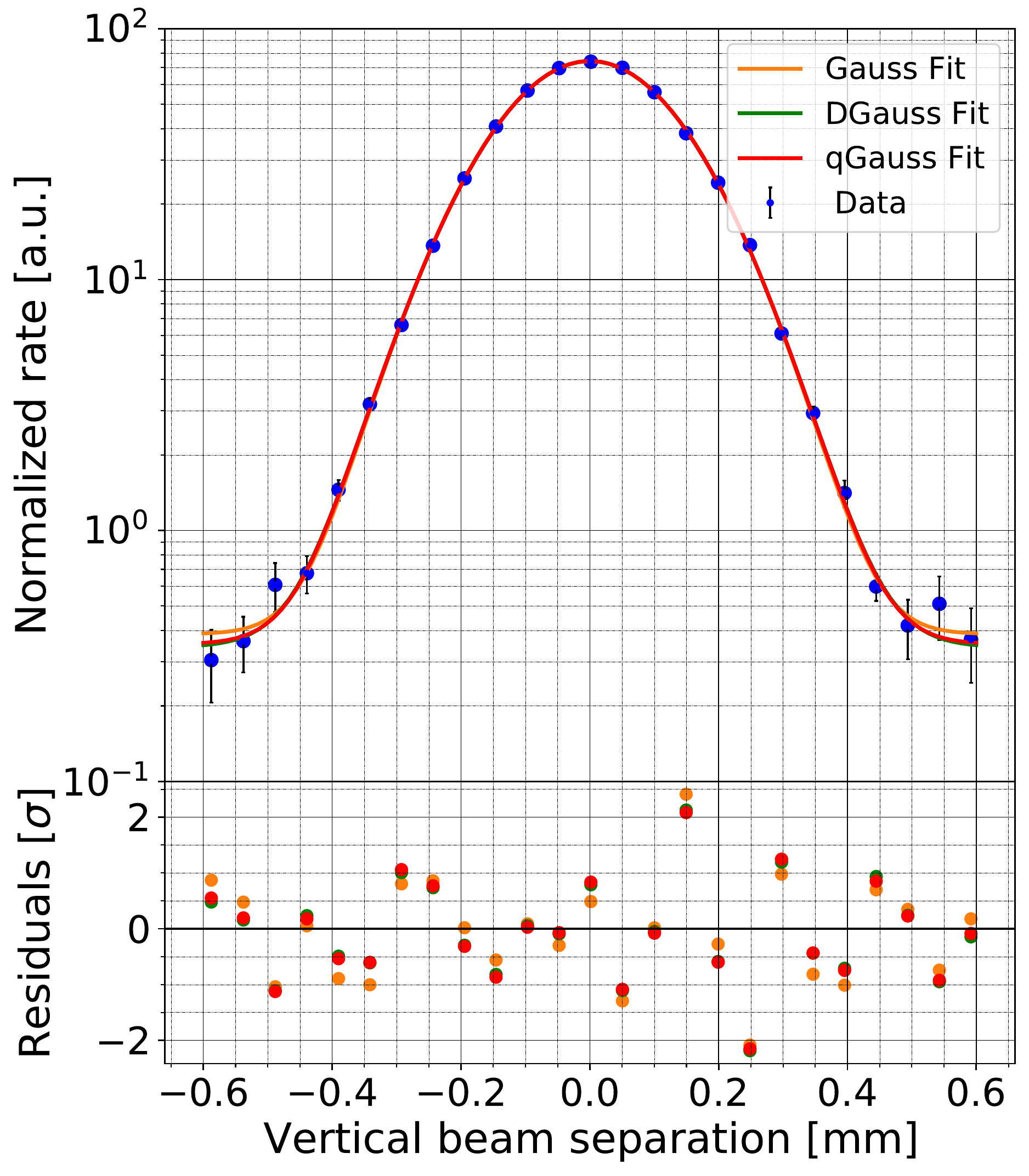}}
		\caption{Fits \eqref{eq:34}, \eqref{eq:35} and \eqref{eq:36} applied to dataset \cite{r10} where the normalized rates are represented as a function of the beam separation in the horizontal $X$ (\ref{subfig:11a}) and vertical $Y$ (\ref{subfig:11b}) directions. The resultant fitting parameters are summarized in Table \ref{tab:4}}
		\label{fig:11}
	\end{figure}

	The Gaussian, double Gaussian, and q-Gaussian fit models are applied to the actual vdM scan dataset, from CMS at the LHC run 2 published in \cite{r10}, to assess the proposed q-Gaussian fit model in comparison with fit models used in \cite{r10}, and to investigate its statistical significance in describing the vdM scan data of actual beams and its ability to predict the overlap integral with higher precision. The previous fit models are rewritten in the general form as:
	\begin{equation}
		f^G(\Delta_u)=A\exp\left( -\frac{(\Delta_u-\mu)^2}{2\Sigma_u^{G^2}}\right)+Const;
		\label{eq:34}
	\end{equation}
	\begin{align}
		f^{DG} (\Delta_u )=&A \Bigg[\epsilon \exp\left( -\frac{(\Delta_u-\mu)^2}{2\sigma_{1u}^{2}}\right)\nonumber\\
		&+(1-\epsilon)\exp\left( -\frac{(\Delta_u-\mu)^2}{2\sigma_{2u}^{2}}\right) \Bigg]+Const;
		\label{eq:35}
	\end{align}
	\begin{equation}
		f_u^{qG}(\Delta_u;q)=A e_q\left( - \frac{(\Delta_u-\mu)^2}{(5-3q)\Sigma_u^{qG^2}}\right)+Const,
		\label{eq:36}
	\end{equation}
	where $Const$ is added to account for the background, and $A$ represents the amplitude of the normalized rates. The fitting of the normalized rates and their fitting residuals are presented in Fig. \ref{fig:11}, where the horizontal $X$ scan in Fig. \ref{subfig:11a} and the vertical $Y$ scan in Fig. \ref{subfig:11b}, the resultant fitting parameters are summarized in Table \ref{tab:4}.
	\begin{table*}
		\caption{The estimated fitting parameters for the fitting of dataset \cite{r10} fitted by Gaussian \eqref{eq:34}, double Gaussian \eqref{eq:35} and q-Gaussian \eqref{eq:36} fit models}
		\label{tab:4}
		\begin{tabular*}{\textwidth}{@{\extracolsep{\fill}}ll@{}}
			\begin{tabular*}{\columnwidth}{@{\extracolsep{\fill}}lccc@{}}
				\hline
				\multicolumn{4}{@{}c}{Dataset \cite{r10}, Horizontal $X$ scan} \\
				\hline
				\multicolumn{3}{@{}c}{Fit parameters} & SD errors \\
				\hline
				Gaussian             	         & $\Sigma_u^G\, [\mu m]$    & $136.34$ & $0.40950$ \\
				             	        		 & $A$   					 & $73.633$ & $0.27912$ \\
				             	       	 		 & $\mu, [\mu m]$   		 & $-0.7451$	& $0.44766$ \\
				             	        		 & $Const$    			   	 & $0.4157$ & $0.03931$ \\
				Double  Gaussian				 & $\epsilon$ 				 & $0.9436$	& $254650$\\
							             	     & $\sigma_{1u}^G\, [\mu m]$ & $136.34$	& $443515$\\
							             	     & $\sigma_{2u}^G\, [\mu m]$ & $136.34$	& $741441$\\
							             	     & $A$   					 & $73.633$ & $0.35316$ \\
							             	     & $\mu, [\mu m]$   		 & $-0.7451$ & $0.50505$ \\
							             	     & $Const$    			   	 & $0.4157$ & $0.05288$ \\
				q-Gaussian						 & $q$						 & $0.9968$	& $0.01151$ \\
										         & $\Sigma_u^{qG}\, [\mu m]$ & $136.22$	& $0.61085$ \\
							             	     & $A$   					 & $73.584$ & $0.33827$ \\
							             	   	 & $\mu, [\mu m]$   	     & $-0.7481$ & $0.45796$ \\
							             	   	 & $Const$    			   	 & $0.42433$ & $0.05076$ \\
				\hline
			\end{tabular*} &
			\begin{tabular*}{\columnwidth}{@{\extracolsep{\fill}}lccc@{}}
				\hline
				\multicolumn{4}{@{}c}{Dataset \cite{r10} Vertical $Y$ scan} \\
				\hline
				\multicolumn{3}{@{}c}{Fit parameters} & SD errors  \\
				\hline
				Gaussian             	         & $\Sigma_u^G \, [\mu m]$    & $132.29$ & $0.42974$ \\
						             	   	 	 & $A$   					 & $73.847$ & $0.31728$ \\
						             	   	 	 & $\mu\, [\mu m]$   		 & $0.3868$	& $0.48266$ \\
						             	   	 	 & $Const$    			   	 & $0.3865$ & $0.03692$ \\
				Double  Gaussian				 & $\epsilon$ 				 & $0.9131$	& $0.64473$\\
						             	   	 	 & $\sigma_{1u}^G \, [\mu m]$ & $129.52$	& $10.8451$\\
						             	   	 	 & $\sigma_{2u}^G\, [\mu m]$ & $159.97$	& $119.629$\\
						             	   	 	 & $A$   					 & $74.084$ & $0.38132$ \\
					             	   	 	   	 & $\mu\, [\mu m]$   		 & $0.3725$ & $0.48775$ \\
						             	   	 	 & $Const$    			   	 & $0.3418$ & $0.08048$ \\
				q-Gaussian						 & $q$						 & $1.0157$	& $0.01227$ \\
						             	   	 	 & $\Sigma_u^{qG}\, [\mu m]$ & $132.87$	& $0.63497$ \\
						             	   	 	 & $A$   					 & $74.104$ & $0.37265$ \\
						             	   	 	 & $\mu\, [\mu m]$   	     & $0.3739$ & $0.47587$ \\
						             	   	 	 & $Const$    			   	 & $0.3521$ & $0.04594$ \\
				\hline
			\end{tabular*}
		\end{tabular*}
	\end{table*}

	It is worth noting that for the horizontal scan, the q-Gaussian fitting predicts slightly underpopulated tails of the scan curve with $q=0.9968$, which is consistent with the resultant double Gaussian fitting parameters, where the fitting parameters show a strong dependence on the initial values with infinite sets of the resultant fitting parameters ($\sigma_1$, $\sigma_2$, $\epsilon$) by which the double Gaussian tends back to a single Gaussian, this affirms that for an underpopulated scan, the double Gaussian is equivalent to a single Gaussian, and it cannot provide further enhancement in precision. For the vertical scan, the q-Gaussian fitting predicts slightly overpopulated tails of the scan curve with $q=1.0157$. The statistical analysis of the fit models and the predicted $\Sigma_u^{fit\ model}$ and $\Omega_u^{fit\ model}$ is summarized in Table \ref{tab:4}, where the RMSE, Adj. $R^2$ and $\chi^2/dof$ are used for the goodness of fit analysis.
	
	\begin{table*}
		\caption{Statistical analysis and the predicted convolved beam size $\Sigma_u^{fit\ model}$ and overlap integral $\Omega_u^{fit\ model}$ from the fitting of dataset \cite{r10} by Gaussian \eqref{eq:34}, double Gaussian \eqref{eq:35} and q-Gaussian \eqref{eq:36} fitting models}
		\label{tab:5}
		\begin{tabular*}{\textwidth}{@{\extracolsep{\fill}}lccccc@{}}
			\hline
			Fitting model & $RMSE$ &  Adj. $R^2$ & $\chi^2/dof$& $\Sigma_u^{fit\ model}\, [\mu m]$ & $\Omega_u^{fit\ model}\, [m^{-1}]$ \\
			\hline
			\multicolumn{6}{@{}c}{Dataset \cite{r10}, Horizontal $X$ scan}\\
			\hline
			Gaussian 	   	 &	$44.0315$ &	$0.999755$ &	$1.04963$ &	$136.349$ &	$2925.885$ \\
			Double Gaussian  &	$44.0315$ &	$0.999729$ &	$1.16012$ &	$136.349$ &	$2925.885$ \\
			q-Gaussian 		 &	$43.9468$ & $0.999743$ &	$1.09788$ &	$136.224$ &	$2925.091$ \\
			\hline
			\multicolumn{6}{@{}c}{Dataset \cite{r10}, Vertical $Y$ scan}\\
			\hline
			Gaussian    	&	$38.5665$ &	$0.999697$ &	$1.03108$ &	$132.2882$ &	$3015.706$ \\
			Double Gaussian &	$37.0326$ &	$0.999691$ &	$1.05077$ &	$132.1726$ &	$3018.343$ \\
			q-Gaussian      &	$37.0722$ & $0.999706$ &	$1.00037$ &	$132.8663$ &	$3020.685$ \\
			\hline
		\end{tabular*}
	\end{table*}
	
	For the horizontal scan, even though the q-Gaussian fit has the lowest RMSE, the Gaussian fit has the closest Adj. $R^2$ and $\chi^2/dof$ to $1$, which is explained by the fact that the predicted $q= 0.9968$  is very close to $1$, this justifies that q-Gaussian represents the best fit. The deviation in the predicted overlap integral by Gaussian and q-Gaussian is $0.027\%$. For the vertical scan, the double Gaussian has the lowest RMSE and the highest $\chi^2/dof$, which shows that double Gaussian could predict the data, but it could not explain the variance well; the q-Gaussian has the closest Adj. $R^2$ and $\chi^2/dof$ to $1$, and it has RMSE very close to the double Gaussian, which shows that the q-Gaussian can be considered the best fit since it can predict the data with high precision and can explain the variance as well. The deviation in the predicted overlap integral by double Gaussian and q-Gaussian is $0.077\%$. 
	
	In general, the fitting results show that the q-Gaussian fit model is a promising base for beam overlap modeling. It can account for the non-Gaussian tails for more precise luminosity calibration, especially for the next upgrades of the current colliders following the increasing demands of collider experiments.
	
\section{Conclusion}
In high-energy colliders, particles in the bunches experience several effects that slightly deviate the particle distributions from the exact Gaussian distribution; hence, using a general distribution function such as q-Gaussian can describe the beam profiles more efficiently. In \cite{r13,r15}, the impact of non-Gaussian tails on absolute luminosity is discussed. In this work, the impact of non-Gaussian tails on the precision of the luminosity calibration by van-der-Meer (vdM) scan is investigated, a vdM scan fit model is proposed, and the impact of the tilt angle in the transverse plane of the colliding bunches is considered where the non-factorization effect is observed.

The overlap integral of q-Gaussian bunch is modeled. An analytical formula shows the dependency of the overlap integral $\Omega_u^{qG}$ on tail density $q$ is derived, equations \eqref{eq:17} and \eqref{eq:19}. The deviation of the overlap integral of q-Gaussian bunches $\Omega_u^{qG}$ from that of Gaussian bunches $\Omega_u^{G}$ during separation scan is shown in Fig. \ref{fig:3}, the area under consideration is divided into 3 dependency regions, where in region-1 $\Omega_u^{qG\ heavy\  tails}>\Omega_u^G>\Omega_u^{qG\ light\ tails}$, in region-2 \linebreak[4] $\Omega_u^{qG\  heavy\  tails}<\Omega_u^G<\Omega_u^{qG\  light\  tails}$ and in region-3 \linebreak[4] $\Omega_u^{qG\ heavy\ tails}>\Omega_u^G>\Omega_u^{qG\ light\ tails}$. The limits of the deviation of $\Omega_u^{qG}$ from $\Omega_u^{G}$ at different regions are summarized in Table \ref{tab:1} for tail density $q$ in the range $0.8$ to $1.2$. The impact of the non-Gaussian tails on vdM scan curve width is investigated where an analytical formula for the convolved beam size of q-Gaussian bunches $\Sigma_u^{qG}$ is derived, as in equation \eqref{eq:22}. The deviation of vdM scan curve width of q-Gaussian bunches $\Sigma_u^{qG}$ from that of Gaussian $\Sigma_u^{G}$ is shown in Fig. \ref{subfig:4b}, which shows that the non-Gaussian tails lead to a deviation up to ± $4\%$ from that of Gaussian for tail density $q$ in the range $0.8$ to $1.2$. Both new formulas of the overlap integral and convolved beam size tend to that of the Gaussian as the tail density $q$ tends to 1 as in equations \eqref{eq:20} and \eqref{eq:23}, which is in agreement with the tendency of the q-Gaussian distribution to Gaussian for $q$ tends to 1.

The impact of the tilt angle in the transverse plane of the colliding bunch on their overlap integral was investigated. In general, the tilt angle leads to a geometrical reduction in the overlap integral, and this reduction depends on the horizontal and vertical tail densities and the ratio between the horizontal and vertical bunch dimensions. For Gaussian bunches, the tilt angle results in non-factorizable densities,  but the resultant overlap integral \eqref{eq:26} considered in vdM scan as the function of beam separation is factorizable. Therefore the overlap integral can be determined by two separate one-dimensional vdM scans in horizontal and vertical directions with no bias as in equation \eqref{eq:27}. For q-Gaussian bunches, the tilt angle results in non-factorizable densities \eqref{eq:28} and a non-factorizable overlap integral \eqref{eq:29}. The detailed study of resulted non-factorization bias is abroad of the topic of this article and will be published elsewhere
	 
The impact of the non-Gaussian tails on the precision of the vdM scan is estimated by assuming the q-Gaussinity of the colliding bunches since the q-Gaussian distribution is a more realistic approximation to the actual bunch profile. A new vdM scan fit model is proposed based on the q-Gaussian distribution equation \eqref{eq:33}. A toy vdM scan is simulated by assuming two q-Gaussian bunches with equal bunch dimensions and tail densities collide head-on. The resultant scan data was fitted by Gaussian and double Gaussian fit models to investigate their precision when applied to non-Gaussian bunches, where their results are compared to that of the proposed q-Gaussian fit model. The fitting results show that unlike the double Gaussian fit, which can be applied only for separation scans with overpopulated tails, the proposed model can be applied for scans with underpopulated and overpopulated tails. It presents a good description of the scan data; it represents the best fit in terms of root mean square error (RMSE) and Adj. $R^2$ for tail density $q$ in the range $0.8$ to $1.2$, it predicts the overlap integral with high precision with a deviation up to $\pm0.025\%$, as shown in Fig. \ref{subfig:9b}, whereas the double Gaussian fit model can only be used for scans with overpopulated tails with deviation up to $0.037\%$. These results are valid for arbitrary q-Gaussian bunches with equal bunch dimensions and tail densities. The vdM scan dataset \cite{r10} was investigated, and the proposed q-Gaussian model \eqref{eq:36} was applied and compared to single Gaussian and double Gaussian models. It was found that the horizontal $X$ scan has underpopulated tails, the double Gaussian fit model tends to a single Gaussian, and it does not provide any enhancement in precision; on the other hand, the q-Gaussian fit model represents the best fit; and it can describe the tails with higher precision. 
	
In conclusion, the effect of the tails on the overlap integral is getting more critical since the collider machines are constantly upgraded and the requirements for luminosity precision became stronger that demands more accurate account for bunch shape; hence, this shape should be taken into account accurately. The results of this study show that the models based on q-Gaussian represents a good base for vdM scan data analysis that can accounts for the non-Gaussian tails of bunches that have overpopulated or underpopulated tails.

\begin{acknowledgements}
This work was supported by The Ministry of Science and Higher Education of the Russian Federation in part of the Science program (Project № FSWW-2023-0003).
\end{acknowledgements}
\appendix
		
\section{The derivation of the overlap integral of q-Gaussian bunches with equal dimensions and tail densities}\label{app:A}
	\subsection{Light-tailed q-Gaussian bunches} \label{subsec:A1}
	Let's rewrite the overlap integral $\Omega_u^{qG}(\Delta_u;q<1)$ in terms of $r$, where $r=\frac{1}{1-q}$, $r\in \mathbb{R}
	$, and $r\ge0$, equation \eqref{eq:16} becomes
	\begin{align}
			\Omega_u^{qG}(\Delta_u;r)_{q<1}&=\frac{\beta^{qG}}{{C^{qG}}^2}\mathop{\mathlarger{\mathlarger{\int}}}_{u_{1}}^{u_{2}} \left[1-\frac{\beta^{qG}}{r}\left(u-\frac{\Delta_u}{2}\right)^2\right]_{+}^{r}\nonumber\\ 
			&\times \left[1-\frac{\beta^{qG}}{r}\left(u+\frac{\Delta_u}{2}\right)^2\right]_{+}^{r}\, du,
		\label{eq:A1}
	\end{align}
	for $\Delta_u\ge0$, the limits of integration are $\big \{ \sqrt{\frac{r}{\beta^{qG}}}-\frac{\Delta_u}{2},$\linebreak[4] $ -\sqrt{\frac{r}{\beta^{qG}}}+\frac{\Delta_u}{2} \big \} $ and by changing the integral variable from $u$ to $t$ as $u=\sqrt{\frac{r}{\beta^{qG}}} \,t$ and $du=\sqrt{\frac{r}{\beta^{qG}}}\, dt$ equation \eqref{eq:A1} can be written as:
	\begin{align}
			\Omega_u^{qG}(\Delta_u;r)_{q<1,\Delta_u\ge0}&=\frac{\sqrt{r\, \beta^{qG}}}{{C^{qG}}^2}  \mathop{\mathlarger{\mathlarger{\int}}}_{-1+\frac{\Delta_t}{2}}^{1-\frac{\Delta_t}{2}} \left(\left(1-\frac{\Delta_t}{2}\right)^2-t^2\right)^r\nonumber \\ &\times \left(\left(1+\frac{\Delta_t}{2}\right)^2-t^2\right)^r\, dt,
		\label{eq:A2}
	\end{align}
	where $\Delta_t=\sqrt{\beta^{qG}/r} \Delta_u$. With further mathematical manipulations, it gives 
	\begin{equation*}
		\begin{split}
			\Omega_u^{qG}&(\Delta_t;r)_{q<1,\Delta_u\ge0}=2\frac{\sqrt{r\, \beta^{qG}}}{{C^{qG}}^2} \left(1-\frac{\Delta_t^2}{4}\right)^{2r} \\ &\times \mathop{\mathlarger{\mathlarger{\int}}}_{0}^{1-\frac{\Delta_t}{2}} \left(1-\frac{t^2}{\left(1-\frac{\Delta_t}{2}\right)^2}\right)^r \left(1-\frac{t^2}{\left(1+\frac{\Delta_t}{2}\right)^2}\right)^r \, dt,
		\end{split}
	\end{equation*}
	changing the integration variable once more from $t$ to $t'$ as $t=(1-\frac{\Delta_t}{2})\,\sqrt{t'}$and $dt=\left(1-\frac{\Delta_t}{2}\right)\,\frac{\sqrt{t'}}{2}\, dt$, we obtain
	\begin{equation*}
		\begin{split}
			\Omega_u^{qG}(\Delta_t;r)&_{q<1,\Delta_u\ge0}=\frac{\sqrt{r	\beta^{qG}}}{{C^{qG}}^2} \left(1-\frac{\Delta_t^2}{4}\right)^{2r} \\ &\times \mathop{\mathlarger{\mathlarger{\int}}}_{0}^{1}\frac{\left(1-{t'}^2\right)^2}{\sqrt{t'}}	\left(1-\frac{\left(1-\frac{\Delta_t}{2}\right)^2}{\left(1+\frac{\Delta_t}{2}\right)^2}t^2
		\right)^r \, dt'.
		\end{split}
	\end{equation*}
	Following the integral form of the Gaussian hypergeometric function $\ _2F_1$ \cite{r25}:
	\begin{equation*}
	_2F_1(a,b;c;z)=Beta(b,c-b)\int_0^{1}\frac{t^{b-1}(1-t)^{c-b-1}}{(1-tz)^a}\, dt,
	\end{equation*}
	the closed form of equation \eqref{eq:A2} is found as
	\begin{align}
			&\Omega_u^{qG}(\Delta_t;r)_{q<1,\Delta_u\ge0}=\frac{\sqrt{r\, \beta^{qG}}}{{C^{qG}}^2}
			\left(1-\frac{\Delta_t}{2}\right) 	\left(1-\frac{\Delta_t^2}{4}\right)^{2r} \nonumber \\ & \times Beta\left(\frac{1}{2},r+1\right)\, \ _2F_1\left(-r,\frac{1}{2};r+\frac{3}{2};\frac{\left(1-\frac{\Delta_t}{2}\right)^2}{\left(1+\frac{\Delta_t}{2}\right)^2}\right).
		\label{eq:A3}
	\end{align}
	by rewriting the equation \eqref{eq:A3} in terms of $q$ and $\Delta_u$, equation \eqref{eq:17} is obtained. Similarly, for $\Delta_u<0$, following the previous steps, equation \eqref{eq:17} can be obtained.
	
	\subsection{Heavy-tailed q-Gaussian bunches} \label{subsec:A2}
	Let's rewrite the overlap integral $\Omega_u^{qG}(\Delta_u;q>1)$ in terms of $r$, where $r=\frac{1}{1-q}$, $r\in \mathbb{R} $, and $r>3/2$, equation \eqref{eq:A1} becomes
	\begin{align}
		\Omega_u^{qG}(\Delta_u;r)_{q>1}&=\frac{\beta^{qG}}{{C^{qG}}^2}\mathop{\mathlarger{\mathlarger{\int}}}_{-\infty}^{\infty} \left[1+\frac{\beta^{qG}}{r}\left(u-\frac{\Delta_u}{2}\right)^2\right)^{-r}\nonumber\\ 
		&\times \left(1+\frac{\beta^{qG}}{r}\left(u+\frac{\Delta_u}{2}\right)^2\right)^{-r}\, du,
		\label{eq:A4}
	\end{align}
	since the two $\Delta_u/2$ displacements of bunches in opposite directions are equivalent to one displacement of one bunch by $\Delta_u$, it gives
	\begin{equation*}
		\begin{split}
			\Omega_u^{qG}(\Delta_u;r)_{q>1}=\frac{\beta^{qG}}{{C^{qG}}^2}\mathop{\mathlarger{\mathlarger{\int}}}_{-\infty}^{\infty} \left(1+\frac{\beta^{qG}}{r}\left(u-\Delta_u\right)^2\right)^{-r}\\ 
			\times \left(1+\frac{\beta^{qG}}{r}u^2\right)^{-r}\, du,
		\end{split}
	\end{equation*}
	Changing the integration variable from $u$ to $t$ as $u=\sqrt{\frac{r}{\beta^{qG}}}\, t$ and $du=\sqrt{\frac{r}{\beta^{qG}}}\, dt$ gives
	\begin{equation*}
		\begin{split}
			\Omega_u^{qG}(\Delta_t&;r)_{q>1}\\&=\frac{\sqrt{r\, \beta^{qG}}}{{C^{qG}}^2}\int_{-\infty}^{\infty} \left(1+\left(t-\Delta_t\right)^2\right)^{-r}
			\left(1+t^2\right)^{-r}\, dt,
		\end{split}
	\end{equation*}
	where $\Delta_t=\sqrt{\frac{\beta^{qG}}{r}}\, \Delta_u$, Using Taylor series expansion for $\big(1+(t-\Delta_t)^2\big)^{-r}$ around $\Delta_t=0$ yields
	\begin{equation*}
		\begin{split}
			\Omega_u^{qG}(\Delta_t;r)_{q>1}&=\frac{\sqrt{r\, \beta^{qG}}}{{C^{qG}}^2} \mathop{\mathlarger{\mathlarger{\int}}}_{-\infty}^{\infty}
			\left(1+t^2\right)^{-r} \\&\times \left[\sum_{i=0}^{\infty} \frac{d^i}{d \Delta^i_t} \left(1+\left(t-\Delta_t\right)^2\right)^{-r}\bigg| _{\Delta_t\to 0} \frac{\Delta_t^i}{i!} \right]
			\, dt,
		\end{split}
	\end{equation*}
	and since $ \frac{d^i}{d \Delta^i_t} \left(1+\left(t-\Delta_t\right)^2\right)^{-r}\bigg| _{\Delta_t\to 0} = \frac{d^i}{d t^i} \left(1+t^2\right)^{-r}$ and by applying Fubini's theorem, we get
	\begin{equation*}
		\begin{split}
			\Omega_u^{qG}(\Delta_t;r)_{q>1}&=\frac{\sqrt{r\, \beta^{qG}}}{{C^{qG}}^2} \\&\times \sum_{t=0}^{\infty}\left[ \int^\infty_{-\infty} (1+t^2)^{-r}\frac{d^i}{d t^i}\left(1+t^2\right)^{-r} \, dt\right]
			\frac{\Delta_t^i}{i!},
		\end{split}
	\end{equation*} 
	since $(1+t^2)^{-r}\frac{d^i}{d t^i}\left(1+t^2\right)^{-r}$ is an even function for the even derivatives and an odd function for the odd derivatives; combined with the fact of the symmetric integration interval, the integration vanishes for odd $i$’s, and we get 
	\begin{equation*}
		\begin{split}
			\Omega_u^{qG}&(\Delta_t;r)_{q>1}=\frac{\sqrt{r\, \beta^{qG}}}{{C^{qG}}^2} \\&\times \sum_{i=0,\,\, i\in Evens}^{\infty}\left[ (-1)^\frac{i}{2}\frac{i!}{(i/2)!}(r)^2_{\frac{i}{2}}\frac{\Gamma({\frac{1}{2})\Gamma({2r-\frac{1+i}{2}})}}{\Gamma({2r+i})}\right]
			\frac{\Delta_t^i}{i!},
		\end{split}
	\end{equation*} 
	where $(\,)_i$ Is the Pochhammer symbol for rising factorial, changing the summation variable from $i$ to $k$ as $i=2k$ and by further mathematical manipulations, we obtain
	\begin{equation*}
		\begin{split}
		\Omega&_u^{qG}(\Delta_t;r)_{q>1}\\&=\frac{\sqrt{r\,\beta^{qG}}}{{C^{qG}}^2} \frac{\Gamma(\frac{1}{2})\Gamma(2r-\frac{1}{2})}{\Gamma(2r}\sum_{k=0}^{\infty} \frac{(r)_k \left(2r-\frac{1}{2}\right)_k}{k!\left(r+\frac{1}{2}\right)_k}\left(-\frac{\Delta_t^2}{4}\right)^k,
		\end{split}
	\end{equation*} 
	Following the definition of the Gaussian hypergeometric function $\ _2F_1$ \cite{r25}:
	\begin{equation*}	
		\ _2F_1(a,b;c;z)=\sum_{k=0}^{\infty}\frac{(a)_k(b)_k}{c_(k)}\frac{z^k}{k!},
	\end{equation*} 
	the closed form of equation \eqref{eq:A4} is obtained as
	\begin{align}
			\Omega_u^{qG}(\Delta_t;r)_{q>1}&=\frac{\sqrt{r\,\beta^{qG}}}{{C^{qG}}^2} Beta\left(\frac{1}{2},2r-\frac{1}{2}\right) \nonumber \\&\times _2F_1\left(r,2r-\frac{1}{2},r+\frac{1}{2};-\beta^{qG}\frac{\Delta_u^2}{4r}\right),
	\label{eq:A5}
	\end{align} 
	by rewriting the equation \eqref{eq:A5} in terms of $q$ and $\Delta_u$, equation \eqref{eq:19} is obtained.   

\section{The tendency of the overlap integral of q-Gaussian bunches to that of Gaussian at the limit of tail density $q$ tends to $1$}\label{app:B}
	\subsection{Light-tailed q-Gaussian bunches} \label{subsec:B1}
	Starting from equation \eqref{eq:A3}, substituting with $C^{qG}$ and $\beta^{qG}$ from equations \eqref{eq:10} and \eqref{eq:11} in terms of $r=\frac{1}{1-q}$, this yields
	\begin{equation*}
		\begin{split}
				\Omega_u^{qG}(\Delta_u;r&)_{q<1}=\frac{1}{Beta\left(\frac{1}{2},r+1\right)\sqrt{2r+3}\,\sigma_u^{qG}}\\
			&\times\left( 1-\sqrt{\frac{\Delta_u^2}{(2r+3){\sigma_u^{qG}}^2}}\right) \left( 1-{\frac{\Delta_u^2}{(2r+3){\sigma_u^{qG}}^2}}\right)^{2r} \\
			&\times\ _2F_1\left(-r,\frac{1}{2};r+\frac{2}{3};\frac{\left( 1-\sqrt{\frac{\Delta_u^2}{(2r+3){\sigma_u^{qG}}^2}}\right)^2}{\left( 1+\sqrt{\frac{\Delta_u^2}{(2r+3){\sigma_u^{qG}}^2}}\right)^2}\right).
		\end{split}
	\end{equation*}
	Since $\lim_{q\to1} \left( \Omega^{qG}_u(\Delta_u;q<1)\right)$ is equivalent to \linebreak[4] $\lim_{r\to\infty} \left( \Omega^{qG}_u(\Delta_u;r)_{q<1}\right)$ and by applying Stirling approximation for Gamma function, with $\ _2F_1 $ definition we get
	\begin{equation*}
		\begin{split}
			\lim_{r\to\infty} \big( \Omega^{qG}_u&(\Delta_u;r)_{q<1}\big)\\&=\frac{1}{\sqrt{\pi\, e}\, \sigma_u^{qG}}\lim_{r\to \infty} 
			\mathop{\mathlarger{\mathlarger{{\mathlarger{\Bigg[}}}}}
			\left(1+\frac{1}{2r}\right)^{r+1} \left(2+\frac{3}{r}\right)^{-\frac{1}{2}}\\&\times
			\left( 1-\sqrt{\frac{\Delta_u^2}{(2r+3){\sigma_u^{qG}}^2}}\right)
			\left( 1-{\frac{\Delta_u^2}{(2r+3){\sigma_u^{qG}}^2}}\right)^{2r}\\&\times
			\sum_{k=0}^{\infty}\frac{\left(\frac{1}{2}\right)_k(-r)_k}{\left(r+\frac{3}{2}\right)_k k!}
			\left(\frac{\left( 1-\sqrt{\frac{\Delta_u^2}{(2r+3){\sigma_u^{qG}}^2}}\right)}{\left( 1+\sqrt{\frac{\Delta_u^2}{(2r+3){\sigma_u^{qG}}}}\right)^2}\right)^{2k}
			\mathop{\mathlarger{\mathlarger{\mathlarger{{\Bigg]}}}}},
		\end{split}
	\end{equation*}
	by applying the limit, we get
	\begin{align}
			\lim_{r\to\infty} \big(& \Omega^{qG}_u(\Delta_u;r)_{q<1}\big)\nonumber\\
			&=\frac{1}{\sqrt{\pi\, e}\, \sigma_u^{qG}} \sqrt{\frac{e}{2}}\exp\left(-\frac{\Delta_u^2}{4{\sigma_u^{qG}}^2}\right)\sum_{k=0}^{\infty}\left(\frac{1}{2}\right)_k\frac{(-1)^k}{k!}\nonumber\\
			&=\frac{1}{2 \sqrt{\pi}\, \sigma_u^{qG}} \exp\left(-\frac{\Delta_u^2}{4{\sigma_u^{qG}}^2}\right)
		\label{eq:B1}
	\end{align}

	\subsection{Heavy-tailed q-Gaussian bunches} \label{subsec:B2}
	Starting from equation \eqref{eq:A5}, substituting with $C^{qG}$ and $\beta^{qG}$ from equations \eqref{eq:10} and \eqref{eq:11} in terms of $r=\frac{1}{q-1}$, this yields
	\begin{equation*}
		\begin{split}
		\Omega_u^{qG}(\Delta_t;r)_{q>1}&=\frac{Beta\left(\frac{1}{2},2r-\frac{1}{2}\right)}{\sqrt{2r-3}\, \sigma_u^{qG}Beta\left(\frac{1}{2},r-\frac{1}{2}\right)}\\& \times
			\, _2F_1\left(r,2r-\frac{1}{2};r+\frac{1}{2};-\frac{\Delta_u^2}{4(2r-3){\sigma_u^{qG}}}^2 \right).
		\end{split}
	\end{equation*}
	Since $\lim_{q\to1} \left( \Omega^{qG}_u(\Delta_u;q>1)\right)$ is equivalent to \linebreak[4] $\lim_{r\to\infty} \left( \Omega^{qG}_u(\Delta_u;r)_{q>1}\right)$ and by applying Stirling approximation for Gamma function, with $\ _2F_1 $ definition we get
	\begin{equation*}
		\begin{split}
			\lim_{r\to\infty} \big( \Omega^{qG}_u&(\Delta_u;r)_{q>1}\big)\\&=\frac{1}{\sqrt{2\pi\, e}\, \sigma_u^{qG}}\lim_{r\to \infty}
			\mathop{\mathlarger{{{{\Bigg[}}}}}
			\left(2-\frac{3}{r}\right)^{\frac{-1}{2}} \left(1-\frac{3}{2r}\right)^{3-2r}\\&\times \left(1-\frac{1}{r}\right)^{2r-1}
			\left(1-\frac{3}{4r}\right)^{2r-1} \left(1-\frac{1}{2r}\right)^{2r-\frac{1}{2}}\\&\times
			\sum_{k=0}^{\infty}\frac{\left(r\right)_k(2r-\frac{1}{2})_k}{r^k\left(r+\frac{1}{2}\right)_k k!}
			\left(-\frac{\Delta_u^2}{4\left(2-\frac{3}{r}\right){\sigma_u^{qG}}^2}\right)^{k}
			\mathop{\mathlarger{{{{\Bigg]}}}}},
		\end{split}
	\end{equation*}
	by applying the limit, we get
	\begin{align}
			\lim_{r\to\infty} \big( \Omega^{qG}_u&(\Delta_u;r)_{q>1}\big)\nonumber\\&
			=\frac{1}{\sqrt{2\pi\, e}\, \sigma_u^{qG}} \frac{e^3 e}{\sqrt{2}e^2 e^{\frac{3}{2}}}\sum_{k=0}^{\infty}\frac{1}{k!}\left(-\frac{\Delta_u^2}{4{\sigma_u^{qG}}^2}\right)\nonumber\\&
			=\frac{1}{2\sqrt{\pi}\, \sigma_u^{qG}}\exp\left(-\frac{\Delta_u^2}{4{\sigma_u^{qG}}^2}\right).
		\label{eq:B2}
	\end{align}
	By comparing the evaluated limits at equations \eqref{eq:B1} and \eqref{eq:B2} with the overlap integral of the Gaussian bunches in equation \eqref{eq:14}, equations \eqref{eq:20} is obtained. 
	
\section{Convolved beam size of q-Gaussian bunches with different bunch dimensions}\label{app:C}
Following equation \eqref{eq:21}, the maximum overlap integral of two light-tailed q-Gaussian beams with densities $q_1=q_2=q$ and has arbitrary bunch dimensions $\sigma_1u^{qG}$ and $\sigma_2u^{qG}$ at zero separation is given by
\begin{equation}
	\begin{split}
		\Omega_u^{qG}(0;q)=\frac{\sqrt{\beta_1^{qG}\beta_2^{qG}}}{{C^{qG}}^2}\int e_q(-\beta_1^{qG}u^2)\, e_q(-\beta_2^{qG}u^2)\, du,
		\label{eq:C1}
	\end{split}
\end{equation}
where $\beta_1^{qG}$ and $\beta_2^{qG}$ can be determined from equation \eqref{eq:11}. The solution of equation \eqref{eq:C1} for light- and heavy-tailed was obtained as 
\begin{equation}
	\begin{split}
		&\Omega_u^{qG}(0;q<1)=\frac{1}{\sqrt{1-q}\sqrt{5-3q}\, \sigma_{max}^{qG} {C^{qG}}^2}\\
		& \times Beta\left(\frac{1}{2},\frac{2-q}{1-q}\right)\ _2F_1 \left( \frac{1}{2},\frac{1}{q-1};\frac{5-3q}{2-2q};\left(\frac{\sigma_{min}^{qG}}{\sigma_{min}^{qG}}\right)^2\right),
		\label{eq:C2}
	\end{split}
\end{equation}
and
\begin{equation}
	\begin{split}
		&\Omega_u^{qG}(0;q>1)=\frac{1}{\sqrt{1-q}\sqrt{5-3q}\, \sigma_{2u}^{qG} {C^{qG}}^2}\\
		& \times Beta\left(\frac{1}{2},\frac{5-q}{2q-1}\right)\ _2F_1 \left( \frac{1}{2},\frac{1}{q-1};\frac{2}{q-1};1-\left(\frac{\sigma_{1u}^{qG}}{\sigma_{2u}^{qG}}\right)^2\right),
		\label{eq:C3}
	\end{split}
\end{equation}
respectively, where $\sigma_{min}^{qG}=min\left(\sigma_{1u}^{qG},\sigma_{2u}^{qG}\right)$ and $\sigma_{max}^{qG}=$\linebreak[4] $max\left(\sigma_{1u}^{qG},\sigma_{2u}^{qG}\right)$. Then the convolved beam size is obtained by substituting equations \eqref{eq:C2} and \eqref{eq:C3} into equation \eqref{eq:21} as:
\begin{align}
		\Sigma_u^{qG}(0;q<1)=\frac{\sigma_{max}^{qG}}{_2F_1 \left( \frac{1}{2},\frac{1}{q-1};\frac{5-3q}{2-2q};\left(\frac{\sigma_{min}^{qG}}{\sigma_{min}^{qG}}\right)^2\right)},
		\label{eq:C4}
\end{align}
and
\begin{align}
		\Sigma_u^{qG}(0;q>1)&=\frac{\beta\left(\frac{1}{2},\frac{3-q}{2q-1}\right)}{\beta\left(\frac{1}{2},\frac{5-q}{2q-1}\right)}\nonumber\\
		&\times \frac{\sigma_{2u}^{qG}}{_2F_1 \left( \frac{1}{2},\frac{1}{q-1};\frac{2}{q-1};1-\left(\frac{\sigma_{1u}^{qG}}{\sigma_{2u}^{qG}}\right)^2\right)}.
		\label{eq:C5}
\end{align}

\end{document}